\documentclass[journal=jacsat,manuscript=article]{achemso}

\usepackage{xr}
\makeatletter
\newcommand*{\addFileDependency}[1]{
  \typeout{(#1)}
  \@addtofilelist{#1}
  \IfFileExists{#1}{}{\typeout{No file #1.}}
}
\makeatother

\newcommand*{\myexternaldocument}[1]{
    \externaldocument{#1}
    \addFileDependency{#1.tex}
    \addFileDependency{#1.aux}
}

\myexternaldocument{supportinginformation}

\usepackage{chemformula} 
\usepackage[T1]{fontenc} 
\usepackage{gensymb}

\author{Vukan Levajac}
\email{v.levajac@tudelft.nl}
\author{Grzegorz P. Mazur}
\author{Nick van Loo}
\author{Francesco Borsoi}
\affiliation[QuTech]
{QuTech and Kavli Institute of Nanoscience, Delft University of Technology, 2600 GA Delft, The Netherlands}
\author{Ghada Badawy}
\author{Sasa Gazibegovic}
\author{Erik P. A. M. Bakkers}
\affiliation[TUe]
{Department of Applied Physics, Eindhoven University of Technology, 5600 MB Eindhoven, The Netherlands}
\author{Sebastian Heedt}
\author{Leo P. Kouwenhoven}
\affiliation[QuTech]
{QuTech and Kavli Institute of Nanoscience, Delft University of Technology, 2600 GA Delft, The Netherlands}
\author{Ji-Yin Wang}
\email{wangjiyinshu@gmail.com}
\affiliation[QuTech]
{QuTech and Kavli Institute of Nanoscience, Delft University of Technology, 2600 GA Delft, The Netherlands}
\alsoaffiliation[Beijing]
{Beijing Academy of Quantum Information Sciences, 100193 Beijing, China (current address)}

\title[Thinnozzle]
    {Impact of junction length on supercurrent resilience against magnetic field in InSb-Al nanowire Josephson junctions}

\keywords{seconductor nanowire, Josephson junction, magnetic field, resilient supercurrent}

\begin{document}


\begin{abstract}
  Semiconducting nanowire Josephson junctions represent an attractive platform to investigate the anomalous Josephson effect and detect topological superconductivity by studying Josephson supercurrent. However, an external magnetic field generally suppresses the supercurrent through hybrid nanowire junctions and significantly limits the field range in which the supercurrent phenomena can be studied. In this work, we investigate the impact of the length of InSb-Al nanowire Josephson junctions on the supercurrent resilience against magnetic fields. We find that the critical parallel field of the supercurrent can be considerably enhanced by reducing the junction length. Particularly, in $30\,\mathrm{nm}$-long junctions supercurrent can persist up to $1.3\,\mathrm{T}$ parallel field - approaching the critical field of the superconducting film. Furthermore, we embed such short junctions into a superconducting loop and obtain the supercurrent interference at a parallel field of $1\,\mathrm{T}$. Our findings are highly relevant for multiple experiments on hybrid nanowires requiring a magnetic field-resilient supercurrent.   
  \end{abstract}


Semiconducting nanowire Josephson junctions (JJs) are widely used as a versatile platform for studying various physical phenomena that arise in semiconductor-superconductor hybrid systems. Therein, the III-V semiconductors have attracted a particular interest in exploring the anomalous Josephson effect\cite{yokoyama_prb_2014, szombati_natphys_2016, strambini_natnano_2020}{}, topological superconductivity \cite{sanjose_prl_2014, lutchyn_prl_2010, oreg_prl_2010, schrade_prb_2017, schrade_prl_2018} and the Josephson diode effect \cite {chen_prb_2018, turini_arx_2022, wu_nat_2022}{}, due to their strong spin-orbit interaction and large $g$ factor. In such experiments, an indispensible ingredient is the Zeeman energy introduced by an external magnetic field. However, an external magnetic field generally suppresses the supercurrent through a hybrid nanowire JJ - therefore significantly limiting the parameter space for addressing the aforementioned effects in hybrid nanowires. Preserving the supercurrent in hybrid nanowire JJs at high magnetic fields becomes thus critically important. Selecting high critical field superconductors, such as NbTiN \cite{gul_nanolett_2017}{}, Pb \cite{kanne_natnano_2021}, Sn \cite{pendharkar_science_2021}{} or Al doped by Pt \cite{mazur_advmater_2022}, seems to be an option for improving the magnetic field compatibility of the supercurrent. However, none of these material platforms have yielded a supercurrent at high magnetic fields. Moreover, it has been observed that the supercurrent of nanowire JJs generally vanishes at magnetic fields far below the critical field of the superconducting film \cite{heedt_natcomm_2021, zuo_prl_2017}{}. Searching for an alternative way to improve the supercurrent resilience against magnetic field in nanowire JJs is thus needed.

In this work, we have studied InSb-Al nanowire JJs with the junction length $L$ varying from $27\,\mathrm{nm}$ to $160\,\mathrm{nm}$. The junction length has been found to be an essential parameter that determines the supercurrent evolution in a parallel magnetic field, as well as the critical parallel field of the supercurrent. In the long devices ($L \sim 160\,\mathrm{nm}$), the supercurrent is suppressed quickly in a magnetic field and fully vanishes at parallel fields of $\sim 0.7\,\mathrm{T}$. In contrast, the supercurrent in short devices ($L \sim 30\,\mathrm{nm}$) persists up to parallel fields of $\sim 1.3\,\mathrm{T}$, approaching the critical in-plane magnetic field of the Al film ($\sim 1.5\,\mathrm{T}$ \cite{mazur_advmater_2022,heedt_natcomm_2021,borsoi_afm_2021}{}). The  evolution of supercurrent in parallel magnetic field is strongly influenced by the electro-chemical potential in all junctions, however, the resilient supercurrent is present only in the short devices ($L \sim 30\,\mathrm{nm}$). We exploit this property to realise a magnetic field-resilient superconducting quantum interference device (SQUID). At the magnetic field of $1\,\mathrm{T}$ parallel to the SQUID arms, the supercurrent through the device displays the characteristic oscillatory pattern as a function of the magnetic flux through the loop. We expect that our demonstration of magnetic field resilient supercurrent in remarkably short nanowire JJs offers a new approach to improving the field-compatibility of not only SQUIDs but many other hybrid nanowire devices utilizing the Josephson effect at high magnetic field.       

\begin{figure}[!t] 
\centering
\includegraphics[width=0.5\linewidth] {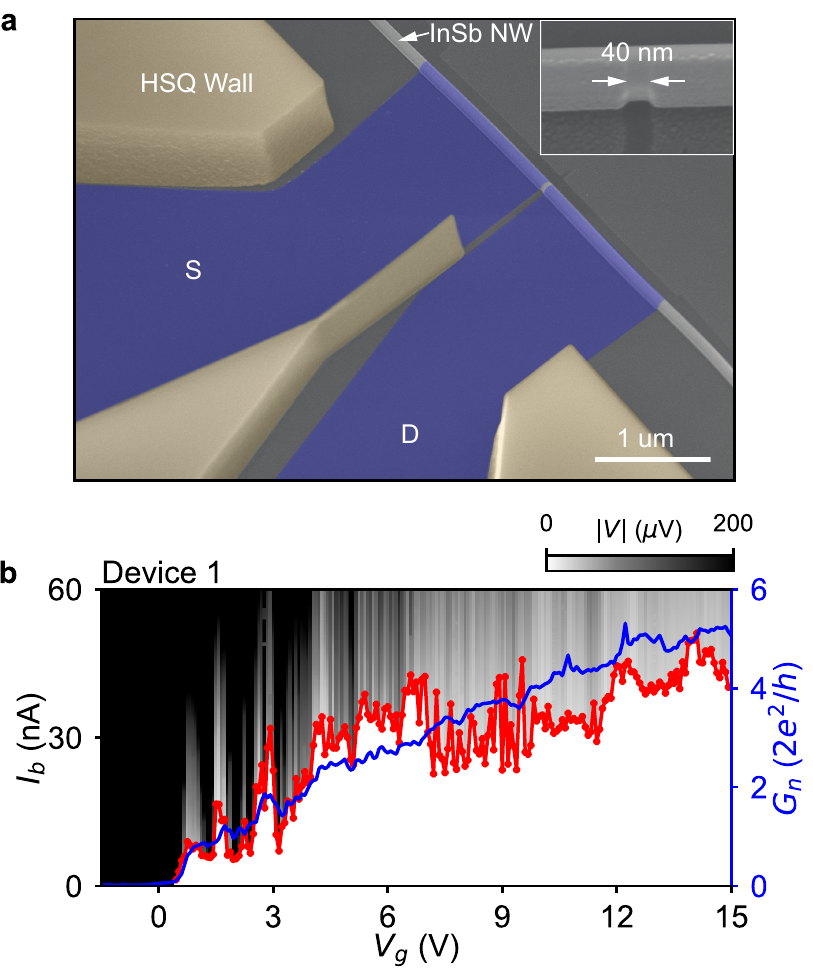}
\caption{\textbf{Basic characterization of a nanowire Josephson junction device}: \textbf{(a)} False-colored SEM image depicting a representative JJ device with a semiconducting InSb junction defined between the source (S) and drain (D) superconducting Al leads (blue). The junction length is determined by the hydrogen silsesquioxane (HSQ) (yellow) shadow-wall structure. A zoom-in at the junction is shown in the inset. The back side of the substrate is used as a global back gate. \textbf{(b)} Zero-field dependence of switching current $I_{sw}$ (red) and normal state conductance $G_n$ (blue) on the back gate voltage $V_{g}$, overlapped onto the $I_{b}$ - $V_{g}$ two-dimensional (2D) map taken for Device 1 (with junction length $L=37\,\mathrm{nm}$).}\label{fig:1}
\end{figure}

The hybrid nanowire JJs are fabricated by the recently developed shadow-wall deposition techniques \cite{heedt_natcomm_2021,borsoi_afm_2021}{}. In Fig. \ref{fig:1}a, a scanning electron microscope (SEM) image of a representative InSb-Al nanowire JJ device is taken at a tilted angle and shown with false colors. Source (S) and drain (D) superconducting leads (blue) are formed via an in-situ angle deposition of Al film after the preparation of a clean and oxide-free InSb nanowire \cite{badawy_nanolett_2019} interface (see the Methods section in the Supporting Information). Pre-patterned dielectric shadow-walls (yellow) selectively define the nanowire sections that are exposed to the Al flux during the deposition. The junction length is determined by the width of the shadow-wall in the vicinity of which the nanowire is deposited. In comparison with the etched dielectric shadow-walls used in recent works \cite{heedt_natcomm_2021,borsoi_afm_2021}{}, here we use lithographically defined shadow-walls which dimensions therefore can be as small as $20\,\mathrm{nm}$. This allows us to precisely control the length of nanowire JJs and to achieve surpassingly short junctions, as shown in the inset SEM image in Fig. \ref{fig:1}a. In this work, nine nanowire JJ devices are presented (Device 1-9) with the junction length $L$ in the range of $27\,\mathrm{nm}-160\,\mathrm{nm}$. The diameter of the nanowires is $\sim 100\,\mathrm{nm}$. An overview of these devices is shown in Fig. S2 in the Supporting Information.

We first characterize the nanowire Josephson junction devices by means of quantum transport at zero magnetic field and $\sim 20\,\mathrm{mK}$ base temperature. The back side of the substrate is used as a back gate and the applied voltage $V_g$ acts globally on the entire nanowire.  In Fig. \ref{fig:1}b we show how the switching current $I_{sw}$ (red) and the normal state conductance $G_n$ (blue) depend on $V_g$ for Device 1. In order to obtain the switching current $I_{sw}$, a four-terminal measurement is employed, where the voltage drop $V$ over the junction is measured while sweeping the current bias $I_b$. $I_{sw}$ is extracted from the $(V,I_b)$ traces as the bias value at which the junction switches to the resistive quasiparticle regime (see the Data analysis section in the Supporting Information for the $I_{sw}$ extraction algorithm). The normal state conductance $G_n$ is obtained in the voltage-bias range $1\,\mathrm{mV}<|V_b|<2\,\mathrm{mV}$ - well above the double value of the induced superconducting gap of the leads ($2\Delta\sim0.5\,\mathrm{meV}$). The conductance measurements from which $G_n$ and $\Delta$ are extracted are shown in Fig. \ref{fig:SVbias} and Fig. \ref{fig:S11} and the procedure for obtaining $G_n$ is explained in the Data analysis section in the Supporting Information. The details on the measurement setups are given in the Measurement section in the Supporting Information. By increasing $V_g$, both $I_{sw}$ and $G_n$, in spite of the fluctuations, become larger as the carrier states in the junction get populated and more subbands contribute to transport. At $V_g=15\,\mathrm{V}$, $G_n$ and $I_{sw}$ reach up to $\sim5\,G_0$ ($G_0=2e^2/h$) and $\sim50\,\mathrm{nA}$, respectively. The remaining nanowire JJs (Device 2-9) show comparable zero-field properties, as shown in Fig. \ref{fig:SVbias} and Fig. \ref{fig:S2} in the Supporting Information. The high tunability of $G_n$ as well as of $I_{sw}$ enables the systematic investigation of the junctions in different electro-chemical potential regimes.  

\begin{figure}[!t] 
\centering
\includegraphics[width=\linewidth] {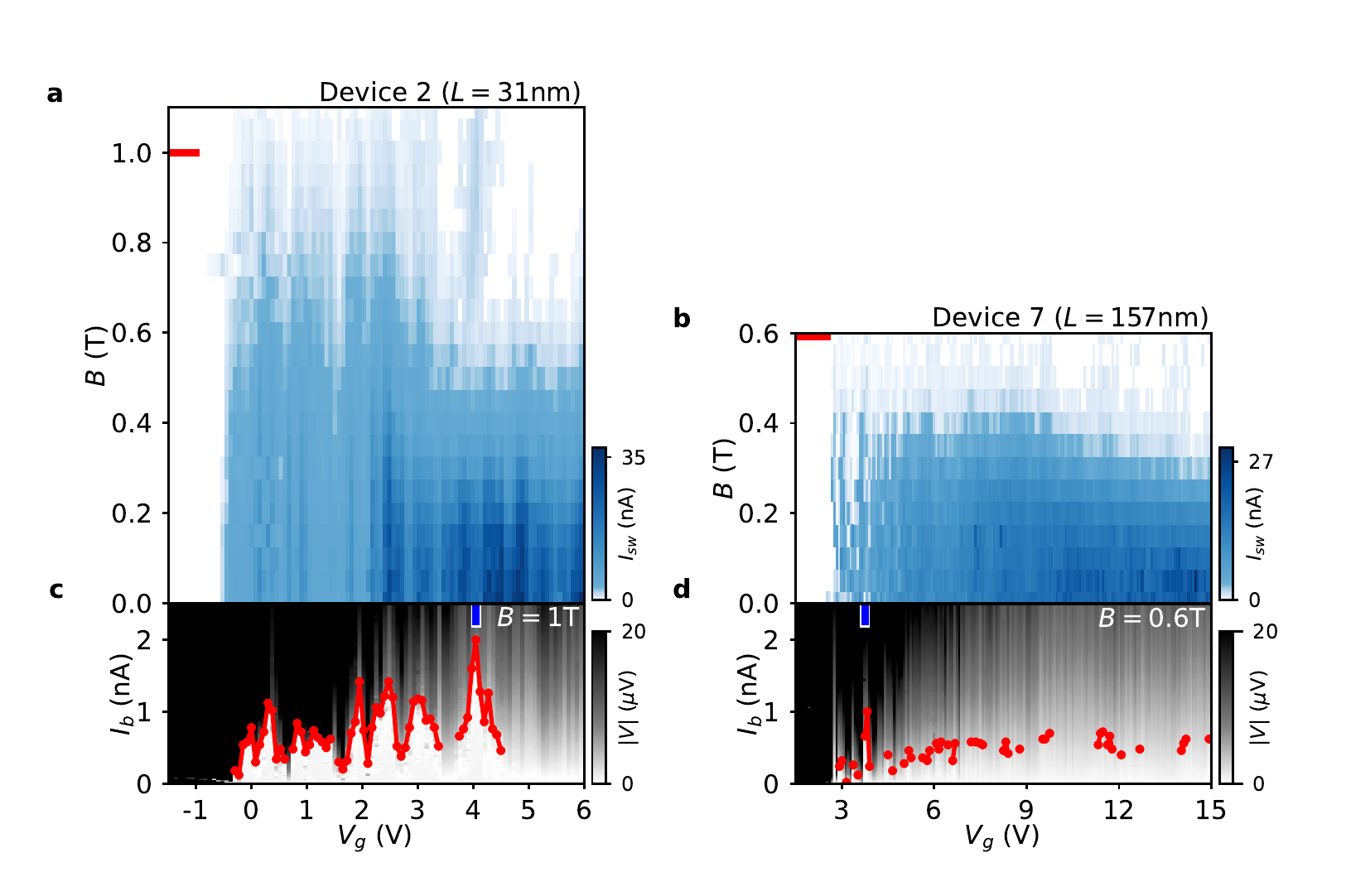}
\caption{\textbf{Dependence of switching current on the gate voltage and parallel magnetic field} for \textbf{(a)} Device 2 ($L=31\,\mathrm{nm}$) and \textbf{(b)} Device 7 ($L=157\,\mathrm{nm}$). Each data point in the $V_g-B$ 2D map in (a) and (b) is extracted from the corresponding $(I_b,V)$ trace as the gate voltage $V_g$ and the parallel magnetic field $B$ are swept. The red markers in (a) and (b) correspond to the magnetic fields $B=1\,\mathrm{T}$ and $B=0.6\,\mathrm{T}$ at which the $I_b-V_g$ 2D maps in \textbf{(c)} and \textbf{(d)} are shown, respectively. In these maps the red traces correspond to the extracted switching current $I_{sw}$. More analogous 2D maps at lower fields are displayed in Fig. S5 in the Supporting Information. The blue markers in (c) and (d) denote the gate settings with enhanced supercurrent.}\label{fig:2}
\end{figure}

Hybrid nanowire JJs have been shown to exhibit a supercurrent evolution in a parallel magnetic field-$B$ that is strongly affected by the electro-chemical potential of the semiconducting junction \cite{zuo_prl_2017}{}. Therefore, when exploring the resilience of switching current in a parallel $B$-field, the electro-chemical potential of a junction has to be taken into account. In the following, the switching current dependence on the back gate voltage $V_g$ and the parallel $B$-field is studied for two JJ devices with significantly different lengths. In Fig. \ref{fig:2}a and \ref{fig:2}b, we show how the switching current $I_{sw}$ evolves with $V_g$ and $B$ for Device 2 ($L=31\,\mathrm{nm}$) and Device 7 ($L=157\,\mathrm{nm}$), respectively. $I_{sw}$ is extracted from the corresponding $(V,I_b)$ traces taken at each setting of $V_g$ and $B$. As shown in Fig. \ref{fig:2}a, the short device shows a remarkable supercurrent resilience with the supercurrent persisting above a parallel field of $1\,\mathrm{T}$. A linecut at $1\,\mathrm{T}$ (red bar) is taken and the corresponding data is shown in Fig. \ref{fig:2}c. The switching current $I_{sw}$ (red trace) continuously persists over a $\sim 3.5\,\mathrm{V}$ interval of $V_g$. As a comparison, $I_{sw}$ drops more rapidly with magnetic field in the long device, as shown in Fig. \ref{fig:2}b. Fig. \ref{fig:2}d shows that at $0.6\,\mathrm{T}$ the supercurrent is barely detectable. Besides this apparent difference, the switching current behaviours in Fig. \ref{fig:2}a and \ref{fig:2}b still show some similarities. Namely, $I_{sw}$ of both devices manifests a better resilience against the magnetic field in an intermediate gate interval between the pinch-off and the fully open regime - $(-0.5,3)\,\mathrm{V}$ interval for the short device and $(4,10)\,\mathrm{V}$ interval for the long device (see Fig. S5 in the Supporting Information). For the gate voltage above these intervals $I_{sw}$ vanishes more rapidly in a magnetic field in both devices. Such behavior at large positive $V_g$ could be explained by a destructive supercurrent interference between multiple modes \cite{zuo_prl_2017} in the open regime. Another explanation could be a reduced semiconductor-superconductor hybridization due to the gate being global and setting a large positive voltage under the superconducting leads \cite{demoor_njphys_2018, shen_prb_2021}{}. This point is addressed in the discussion part following Fig. \ref{fig:4}. An ubiquitous feature in Fig. \ref{fig:2}a and \ref{fig:2}b is that, as the magnetic field is increased, certain intervals in the intermediate gate regime support more resilient supercurrent. In these $V_g$ intervals we define the “resilient gate settings $V_{g,res}$” (blue markers in Fig. \ref{fig:2}c and \ref{fig:2}d) and the supercurrent at such gate settings is examined for all devices in the following paragraph.  

\begin{figure}[!t] 
\centering
\includegraphics[width=\linewidth] {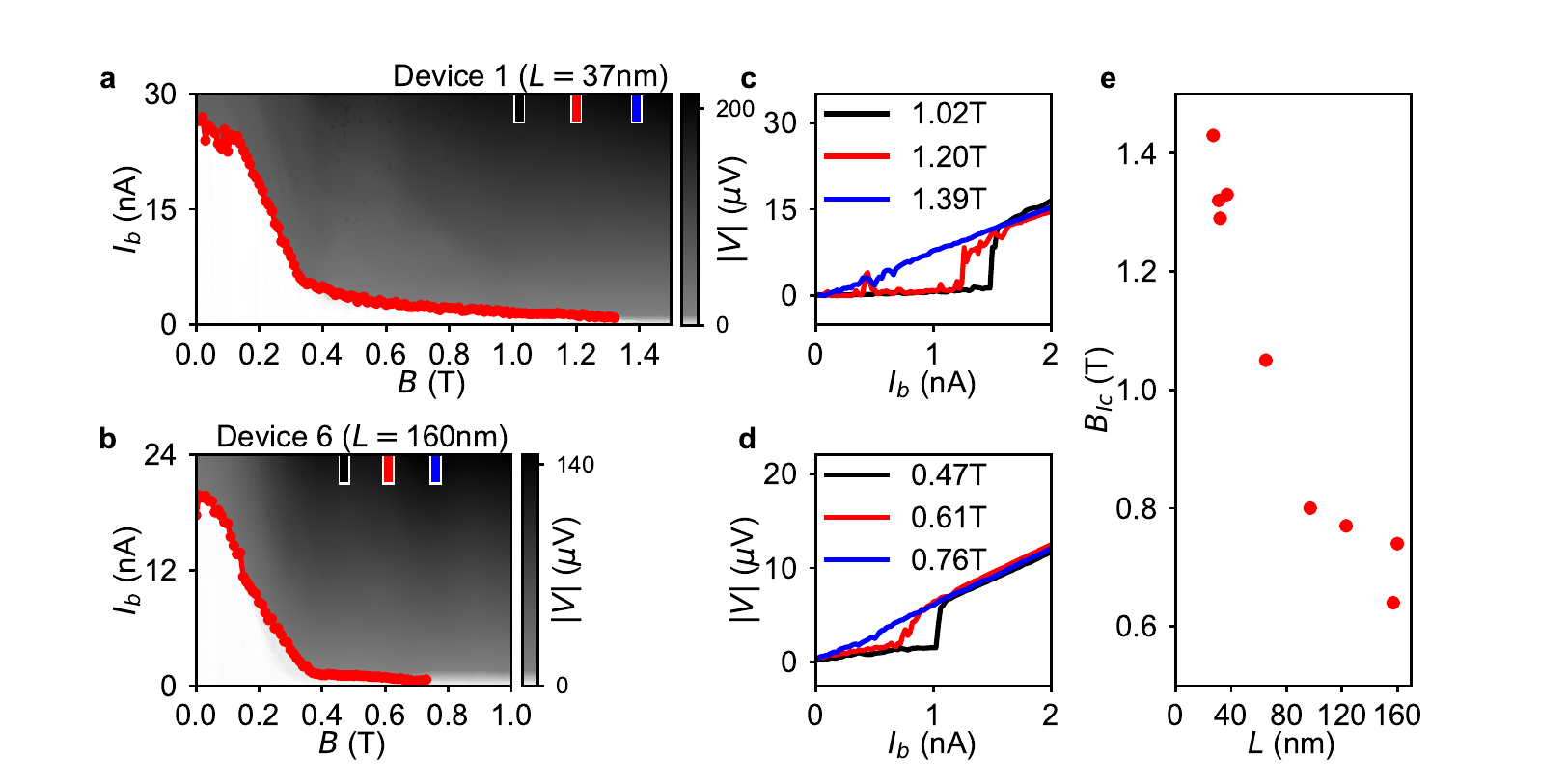}
\caption{\textbf{ Critical parallel magnetic field of switching current}: Dependence of the switching current (red) on $B$ at the resilient gate settings $V_{g,res}$ for \textbf{(a)} Device 1 ($L=37\,\mathrm{nm}$) and \textbf{(b)} Device 6 ($L=160\,\mathrm{nm}$). In each 2D map the extracted switching current $I_{sw}$ up to the critical parallel field is plotted in red. The critical parallel fields of the switching current in (a) and (b) are $B_{Ic}=1.33\,\mathrm{T}$ and $B_{Ic}=0.74\,\mathrm{T}$, respectively. Black, red and blue markers in (a) and (b) have the corresponding linecuts shown in \textbf{(c)} and \textbf{(d)}. In \textbf{(e)} the dependence of the critical parallel field $B_{Ic}$ is plotted for Device 1-9 versus the junction length $L$.}\label{fig:3}
\end{figure}

In Fig. \ref{fig:3} we focus on the supercurrent at the resilient gate settings $V_{g,res}$. For Device 1-7 we determine the $V_{g,res}$ values as shown and described in Fig. S6, while for Device 8-9 we choose  $V_g=15\,\mathrm{V}$ (see the Data selection and reproducibility section in the Supporting Information). Fig. \ref{fig:3}a shows the voltage drop $V$ over the junction as a function of the current bias $I_b$ and the parallel magnetic field $B$ for Device 1 ($L=37\,\mathrm{nm}$). The red dotted line marks the extracted switching current $I_{sw}$ at different  $B$-fields. Three linecuts (black, red and blue) are shown in Fig. \ref{fig:3}c - demonstrating more than $1\,\mathrm{nA}$ supercurrent at the parallel field of $1.2\,\mathrm{T}$. Fig. \ref{fig:3}b and Fig. \ref{fig:3}d show the results for Device 6 ($L=160\,\mathrm{nm}$) obtained at its $V_{g,res}$ setting. From the overlaid red trace it can be seen that the supercurrent vanishes at $\sim0.75\,\mathrm{T}$, as confirmed by the linecuts shown in Fig. \ref{fig:3}d. Analogous measurements of the switching current evolution with parallel field are carried out for all nine devices (see Fig. S7 in the Supporting Information). Finally, these $I_{sw}(B)$ dependences allow for the extraction of the maximal critical parallel magnetic field of switching current $B_{Ic}$ for each Device 1-9. The details of the $B_{Ic}$ extraction are given in the Data analysis section in the Supporting Information. By plotting $B_{Ic}$ versus the junction length $L$ in Fig. \ref{fig:3}e, it can be seen how the junction length influences the measured critical field of the supercurrent. We reproducibly reach the critical fields of $\sim1.3\,\mathrm{T}$ in the sub-$40\,\mathrm{nm}$ junctions while $B_{Ic}$ drops gradually to $\sim0.7\,\mathrm{T}$ in the longest junctions.   

\begin{figure}[!t] 
\centering
\includegraphics[width=0.5\linewidth] {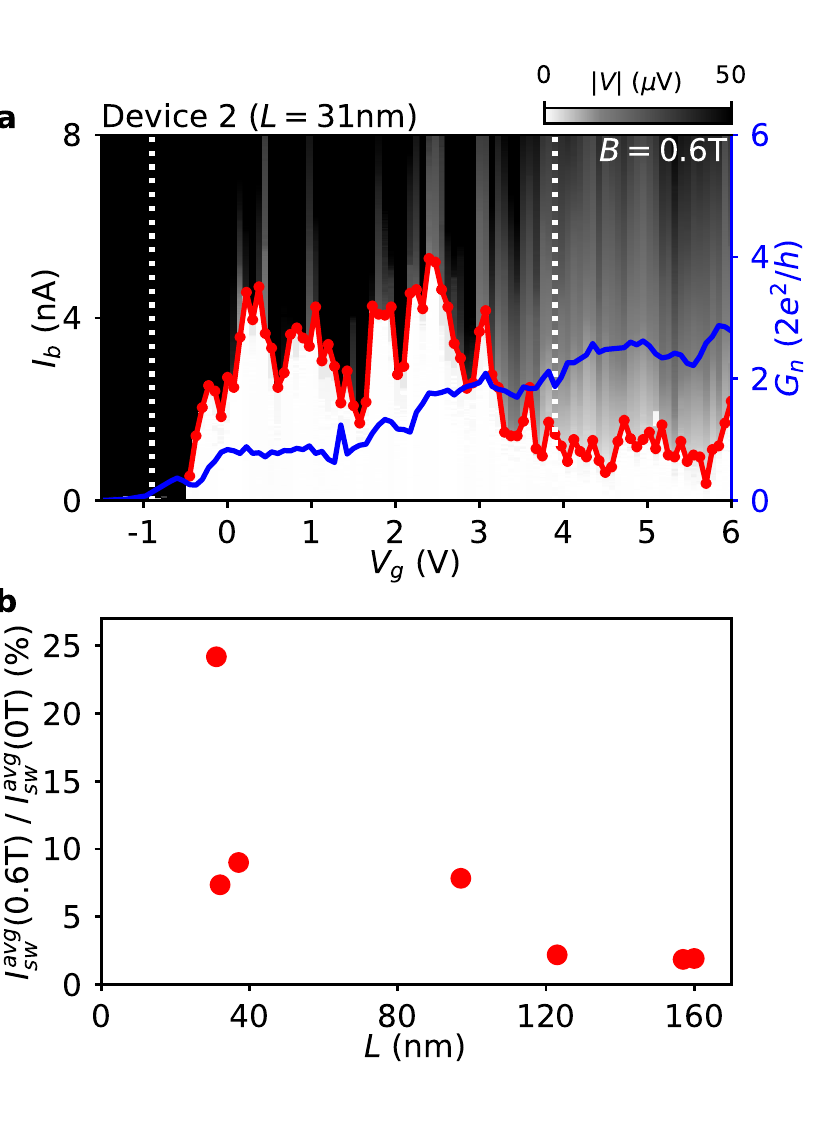}
\caption{\textbf{Resilience of switching current in the junctions tunability ranges:} \textbf{(a)} Dependence of the switching current $I_{sw}$ (red) on the gate voltage $V_g$ at the parallel magnetic field $B=0.6\,\mathrm{T}$ for Device 2 ($L=31\,\mathrm{nm}$). Two white vertical lines mark the gate interval over which the normal state conductance $G_n$ of the device (blue) is tuned from $0.01G_0$ to $2G_0$. In this gate range the switching current is averaged and $I_{sw}^{avg}(0.6\,\mathrm{T})$ value is obtained.  Analogously, from the switching current dependence on $V_g$ at zero-field the average value $I_{sw}^{avg}(0\,\mathrm{T})$ is calculated.  \textbf{(b)} Dependence of the ratio $I_{sw}^{avg}(0.6\,\mathrm{T})/I_{sw}^{avg}(0\,\mathrm{T})$ on the junction length $L$ for Device 1-7. The average values and the corresponding gate ranges of averaging are shown for all devices in Fig. S4 and Fig. S8 in the Supporting Information.}\label{fig:4} 
\end{figure}

As a next step, we evaluate the supercurrent resilience over a broader gate interval - in a range of the electro-chemical potential. As our nanowire JJs are highly tunable, in Fig. \ref{fig:4} their supercurrent resilience against the parallel magnetic field is studied over the gate ranges in which the junctions are in the few mode regimes. Fig. \ref{fig:4}a shows the voltage drop $V$ as a function of the current bias $I_b$ and the gate voltage $V_g$ at the parallel field of $0.6\,\mathrm{T}$ for Device 2 ($L=31\,\mathrm{nm}$), together with the switching current $I_{sw}$ (red trace) and the normal state conductance $G_n$ (blue trace). To quantify the supercurrent resilience, the switching current in Fig. \ref{fig:4}a is averaged in the $V_g$ range corresponding to $0.01G_0<G_n(V_g)<2G_0$ (denoted by the two white dotted vertical lines) and the obtained average switching current is $I_{sw}^{avg}(0.6\,\mathrm{T})=2.73\,\mathrm{nA}$. An analogous averaging is done for the $I_{sw}(V_g)$ dependence measured at zero field and the obtained average switching current at zero field is $I_{sw}^{avg}(0\,\mathrm{T})=11.29\,\mathrm{nA}$ (see Fig. S4 for the zero-field dependence and the average value). By calculating the ratio $I_{sw}^{avg}(0.6\,\mathrm{T})/I_{sw}^{avg}(0\,\mathrm{T})$, it can be inferred that the junction of Device 2 preserves on average $\sim25\,\mathrm{\%}$ of its zero field switching current when the parallel field of $0.6\,\mathrm{T}$ is applied. The identical procedures of calculating the average switching currents and the $I_{sw}^{avg}(0.6\,\mathrm{T})/I_{sw}^{avg}(0\,\mathrm{T})$ ratios are carried out for Device 1-7 (see Fig. S4 and Fig. S8 in the Supporting Information). The dependence of the $I_{sw}^{avg}(0.6\,\mathrm{T})/I_{sw}^{avg}(0\,\mathrm{T})$ on the junction length $L$ is shown in Fig. \ref{fig:4}b. It can be noticed that at finite parallel field the shorter junctions preserve larger fractions of the corresponding zero field supercurrent in the described conductance ranges. Moreover, only negligible fractions of switching current (less than $2\,\mathrm{\%}$) systematically remain in the longer junctions - emphasizing their poor performance when the supercurrent resilience in tunable junctions is of interest. In comparison with Fig. \ref{fig:3}e, the less strong dependence on the junction length can be seen in Fig. \ref{fig:4}b. This can be attributed to particular device-specific shapes of the switching current and the normal state conductance dependences on the gate voltage. We emphasize that the particular shape of the dependence in Fig. \ref{fig:4}b could also vary depending on the choice of the normal conductance tunability range and the subsequently determined gate intervals for averaging. However, the main qualitative features of such dependence would still remain (see Fig. S8 in the Supporting Information).

In Fig. \ref{fig:3} and Fig. \ref{fig:4} two different approaches have been taken when quantifying the supercurrent resilience against magnetic field. Both approaches have led to the same observation - by reducing the junction length supercurrent resilience against magnetic field can be significantly improved. This is a common and reproducible feature of the short JJs in our study. It is observed despite variations in the switching current dependences on the gate voltage or the parallel field (see Fig. S7 and Fig. S8). In order to better understand the interplay between the junction length and the applied gate voltage in determining the supercurrent resilience, we perform two additional measurements that are in detail discussed in the sections describing Fig. S9 and Fig. S10 in the Supporting Information. In the following three paragraphs these measurements are shortly summarized and possible origins of the reported supercurrent resilience in the short JJs are considered at the mesoscopic level.  


Induced superconducting gaps for all nanowire devices and their dependence on the parallel field are studied in Fig. S9 and the corresponding section in the Supporting Information. We find that the measured induced gaps at zero field and the parallel field at which the induced gaps close are influenced not only by the back-gate voltage but by the junction length as well. This is in accordance with the known phenomenon that the proximity effect in nanowire hybrids can be controlled by the electric field \cite{mikkelsen_prx_2018,antipov_prx_2018,demoor_njphys_2018,shen_prb_2021} - that is influenced by both the gate setting and the device geometry. The measurements in Fig. S9 show that reducing the junction length of a hybrid nanowire JJ enhances the proximity effect inside the junction. Consequently, the resilient supercurrent in the short junctions could be attributed to the enhancement of the proximity effect. Possible origins of such enhancement are discussed in more detail in the corresponding section in the Supporting Information.    

Measurements on an additional short JJ device (Device 10, $L=40\,\mathrm{nm}$) are shown in Fig. S10 and are discussed in the corresponding section in the Supporting Information. This device utilizes a bottom gate under the junction and one bottom gate under each superconducting lead. The supercurrent exhibits a nonmonotonic dependence on each of the three bottom gates especially in finite fields. This suggests that the electro-chemical potentials under the superconducting leads and in the junction form a three-dimensional parameter space in which supercurrent is defined. The back gate sweeps in our study correspond to single linecuts in this space and the supercurrent at the resilient gate settings of the back gate may not be the most resilient in the entire parameter space. Importantly, we find in Fig. S10 that applying a positive gate voltage locally under a single superconducting lead does not reduce the superconductor-semiconductor coupling to an extent that systematically limits the resilience of supercurrent. 

Multimode interference is an additional mechanism that could cause the prominent supercurrent dependence on the gate voltage and magnetic field - resulting in the observation of resilient gate settings. Differences between the accumulated phases of different transversal nanowire modes increase with the junction length and the flux applied through the conductive cross-section of a junction \cite{gharavi_prb_2015}. Therefore, destructive supercurrent interference due to large accumulated phase differences could be causing poor supercurrent resilience in long junctions and in open regime in all the junctions - where the conductive section increases due to high positive back gate. Furthermore, in the previous study \cite{zuo_prl_2017} the scattering on disorder was shown to enhance multimode interference. With assuming comparable linear densities of disorder in the junctions in our study, the scattering on disorder would be more prominent in longer junctions and could therefore additionally diminish their supercurrent resilience.  

\begin{figure}[!t] 
\centering
\includegraphics[width=0.5\linewidth] {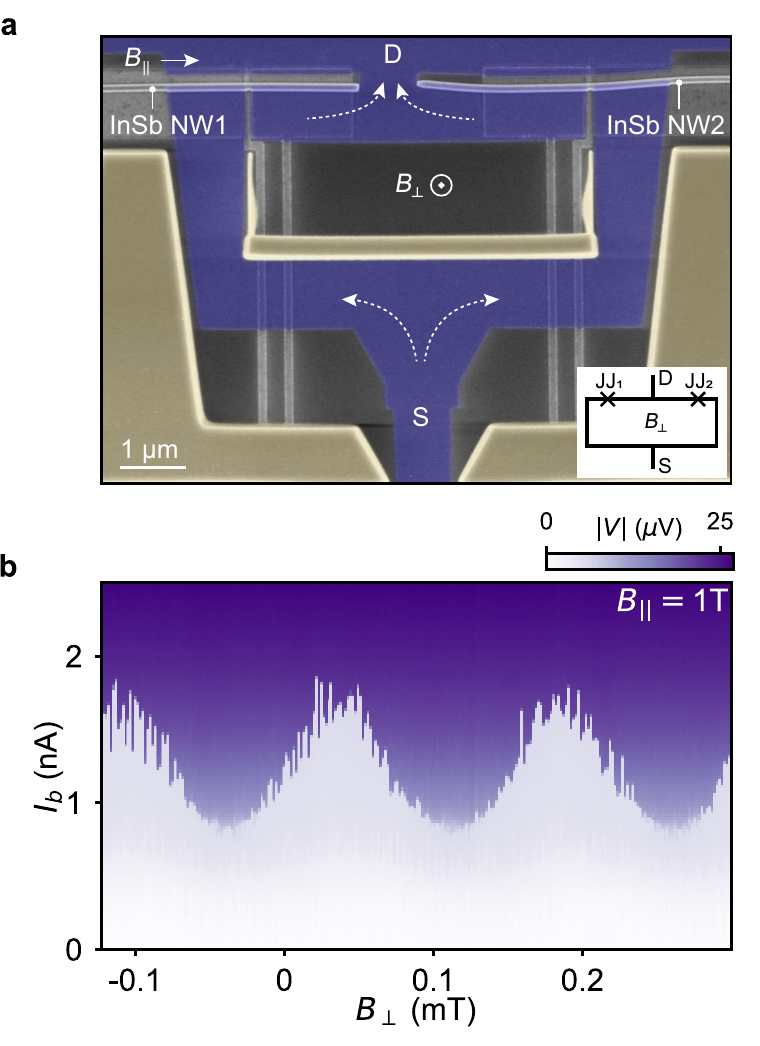}
\caption{\textbf{SQUID operating at a parallel magnetic field of $1\,\mathrm{T}$}: \textbf{(a)} False-colored SEM image of two hybrid $40\,\mathrm{nm}$ long InSb-Al nanowire Josephson junctions defined by the shadow-walls (yellow). The two junctions enclose a superconducting Al (blue) loop in the SQUID architecture. A magnetic field $B_{||}$ is applied along two parallel InSb nanowires hosting the Josephson junctions. A perpendicular out-of-plane magnetic field $B_\perp$ controls the magnetic flux through the superconducting loop between the source (S) and the drain (D). The inset image displays the equivalent device circuit; \textbf{(b)} Current bias measurement on the SQUID at the parallel magnetic field $B_{||}=1\,\mathrm{T}$ shows oscillations of the SQUID switching current as the magnetic flux through the SQUID loop is swept by applying $B_\perp$.}\label{fig:5}
\end{figure}

From the above results, we find that significantly reducing the nanowire JJ length is essential for preserving supercurrents in a high parallel magnetic field. Here, we take a step further and incorporate remarkably short nanowire JJs into the SQUID architecture. Fig. \ref{fig:5}a shows a false-colored SEM of a SQUID consisting of two $40\,\mathrm{nm}$ long hybrid JJs formed in two parallel InSb nanowires. The shadow-wall structure (yellow) is lithographically defined such that after the Al (blue) deposition two JJs enclose the superconducting loop denoted by the white arrows. Since the two arms are parallel, a magnetic field $B_{||}$ can be applied parallel to both JJs while the out-of-plane perpendicular magnetic field $B_{\perp}$ is applied to sweep the flux threading the SQUID loop. Upon applying $B_{||}=1\,\mathrm{T}$, both junctions are independently tuned by the underlying local bottom gates to a finite supercurrent. As shown in Fig. \ref{fig:5}b, the oscillations of the switching current indicate a supercurrent interference persisting despite the high parallel field being applied. In comparison with the previous work on nanowire SQUIDs \citep{wang_sciadv_2022, szombati_natphys_2016}, this observation of supercurrent interference at $B_{||}=1\,\mathrm{T}$ represents a significant improvement of the SQUID field compatibility. The control and the detection of the phase of supercurrent at high magnetic field is of crucial importance for studying various high field related phenomena in hybrid nanowire devices \citep{cxliu_prb_2021, schrade_prl_2018,constantin_prl_2018}.         

In conclusion, we demonstrate that the length of a hybrid nanowire Josephson junction is an essential parameter that determines its supercurrent resilience against magnetic fields. Nanowire JJs with a length of less than $40\,\mathrm{nm}$ can be precisely defined by the shadow-wall angle-deposition technique and are shown to reproducibly preserve supercurrent at parallel magnetic fields exceeding $1.3\,\mathrm{T}$. Superconducting quantum interference device (SQUID) utilizing such  junctions displays supercurrent interference at the parallel field of $1\,\mathrm{T}$. Our study shows that hybrid nanowire Josephson junctions of significantly reduced junction length can be considered as necessary building blocks in various hybrid nanowire devices which exploit Josephson coupling at high magnetic field.      

\section{Author contributions}

J.-Y.W. and V.L. concieved the experiment. J.-Y.W., S.H., G.P.M., N.v.L. and F.B. contributed to the substrate fabrication and/or conducted the superconductor growth. G.B, S.G. and E.P.A.M.B. grew the semiconducting nanowires. V.L. and J.-Y.W. performed the transport measurements. V.L. performed the data analysis. J.-Y.W. and L.P.K.  supervised the project. V.L. and J.-Y.W. wrote the manuscript with inputs from all the other authors.  

\section{Acknowledgements}

The authors thank Chun-Xiao Liu, Michael Wimmer and Anton R. Akhmerov for fruitful discussions. The authors are grateful to Olaf Benningshof and Jason Mensingh for important technical support. This work was financially supported by the Dutch Organization for Scientific Research (NWO), the Foundation for Fundamental Research on Matter (FOM) and Microsoft Corporation Station Q.
      
\section{Conflict of interests}

The authors declare no conflicts of interest. 

\bibliography{references}

\providecommand{\latin}[1]{#1}
\makeatletter
\providecommand{\doi}
  {\begingroup\let\do\@makeother\dospecials
  \catcode`\{=1 \catcode`\}=2 \doi@aux}
\providecommand{\doi@aux}[1]{\endgroup\texttt{#1}}
\makeatother
\providecommand*\mcitethebibliography{\thebibliography}
\csname @ifundefined\endcsname{endmcitethebibliography}
  {\let\endmcitethebibliography\endthebibliography}{}
\begin{mcitethebibliography}{28}
\providecommand*\natexlab[1]{#1}
\providecommand*\mciteSetBstSublistMode[1]{}
\providecommand*\mciteSetBstMaxWidthForm[2]{}
\providecommand*\mciteBstWouldAddEndPuncttrue
  {\def\EndOfBibitem{\unskip.}}
\providecommand*\mciteBstWouldAddEndPunctfalse
  {\let\EndOfBibitem\relax}
\providecommand*\mciteSetBstMidEndSepPunct[3]{}
\providecommand*\mciteSetBstSublistLabelBeginEnd[3]{}
\providecommand*\EndOfBibitem{}
\mciteSetBstSublistMode{f}
\mciteSetBstMaxWidthForm{subitem}{(\alph{mcitesubitemcount})}
\mciteSetBstSublistLabelBeginEnd
  {\mcitemaxwidthsubitemform\space}
  {\relax}
  {\relax}

\bibitem[Yokoyama \latin{et~al.}(2014)Yokoyama, Eto, and
  Nazarov]{yokoyama_prb_2014}
Yokoyama,~T.; Eto,~M.; Nazarov,~Y.~V. Anomalous Josephson effect induced by
  spin-orbit interaction and Zeeman effect in semiconductor nanowires.
  \emph{Phys. Rev. B} \textbf{2014}, \emph{89}, 195407\relax
\mciteBstWouldAddEndPuncttrue
\mciteSetBstMidEndSepPunct{\mcitedefaultmidpunct}
{\mcitedefaultendpunct}{\mcitedefaultseppunct}\relax
\EndOfBibitem
\bibitem[Szombati \latin{et~al.}(2016)Szombati, Nadj-Perge, Car, Plissard,
  Bakkers, and Kouwenhoven]{szombati_natphys_2016}
Szombati,~D.~B.; Nadj-Perge,~S.; Car,~D.; Plissard,~S.~R.; Bakkers,~E. P.
  A.~M.; Kouwenhoven,~L.~P. Josephson $\phi_0$-junction in nanowire quantum
  dots. \emph{Nature Phys.} \textbf{2016}, \emph{12}, 568--572\relax
\mciteBstWouldAddEndPuncttrue
\mciteSetBstMidEndSepPunct{\mcitedefaultmidpunct}
{\mcitedefaultendpunct}{\mcitedefaultseppunct}\relax
\EndOfBibitem
\bibitem[Strambini \latin{et~al.}(2020)Strambini, Iorio, Durante, Citro,
  Sanz-Fern\'{a}ndez, Guarcello, Tokatly, Braggio, Rocci, Ligato, Zannier,
  Sorba, Bergeret, and Giazotto]{strambini_natnano_2020}
Strambini,~E.; Iorio,~A.; Durante,~O.; Citro,~R.; Sanz-Fern\'{a}ndez,~C.;
  Guarcello,~C.; Tokatly,~I.~V.; Braggio,~A.; Rocci,~M.; Ligato,~N.;
  Zannier,~V.; Sorba,~L.; Bergeret,~F.~S.; Giazotto,~F. A Josephson phase
  battery. \emph{Nat. Nanotechnol.} \textbf{2020}, \emph{15}, 656--660\relax
\mciteBstWouldAddEndPuncttrue
\mciteSetBstMidEndSepPunct{\mcitedefaultmidpunct}
{\mcitedefaultendpunct}{\mcitedefaultseppunct}\relax
\EndOfBibitem
\bibitem[San-Jose \latin{et~al.}(2014)San-Jose, Prada, and
  Aguado]{sanjose_prl_2014}
San-Jose,~P.; Prada,~E.; Aguado,~R. Mapping the topological phase diagram of
  multiband semiconductors with supercurrents. \emph{Phys. Rev. Lett.}
  \textbf{2014}, \emph{112}, 137001\relax
\mciteBstWouldAddEndPuncttrue
\mciteSetBstMidEndSepPunct{\mcitedefaultmidpunct}
{\mcitedefaultendpunct}{\mcitedefaultseppunct}\relax
\EndOfBibitem
\bibitem[Lutchyn \latin{et~al.}(2010)Lutchyn, Sau, and Sarma]{lutchyn_prl_2010}
Lutchyn,~R.~M.; Sau,~J.~D.; Sarma,~S.~D. Majorana fermions and a topological
  phase transition in semiconductor-superconductor heterostructures.
  \emph{Phys. Rev. Lett.} \textbf{2010}, \emph{105}, 077001\relax
\mciteBstWouldAddEndPuncttrue
\mciteSetBstMidEndSepPunct{\mcitedefaultmidpunct}
{\mcitedefaultendpunct}{\mcitedefaultseppunct}\relax
\EndOfBibitem
\bibitem[Oreg \latin{et~al.}(2010)Oreg, Refael, and von Oppen]{oreg_prl_2010}
Oreg,~Y.; Refael,~G.; von Oppen,~F. Helical liquids and Majorana bound states
  in quantum wires. \emph{Phys. Rev. Lett.} \textbf{2010}, \emph{105},
  177002\relax
\mciteBstWouldAddEndPuncttrue
\mciteSetBstMidEndSepPunct{\mcitedefaultmidpunct}
{\mcitedefaultendpunct}{\mcitedefaultseppunct}\relax
\EndOfBibitem
\bibitem[Schrade \latin{et~al.}(2017)Schrade, Hoffman, and
  Loss]{schrade_prb_2017}
Schrade,~C.; Hoffman,~S.; Loss,~D. Detecting topological superconductivity with
  $\phi_0$-Josephson junctions. \emph{Phys. Rev. B} \textbf{2017}, \emph{95},
  195421\relax
\mciteBstWouldAddEndPuncttrue
\mciteSetBstMidEndSepPunct{\mcitedefaultmidpunct}
{\mcitedefaultendpunct}{\mcitedefaultseppunct}\relax
\EndOfBibitem
\bibitem[Schrade and Fu(2018)Schrade, and Fu]{schrade_prl_2018}
Schrade,~C.; Fu,~L. Majorana Superconducting Qubit. \emph{Phys. Rev. Lett.}
  \textbf{2018}, \emph{121}, 267002\relax
\mciteBstWouldAddEndPuncttrue
\mciteSetBstMidEndSepPunct{\mcitedefaultmidpunct}
{\mcitedefaultendpunct}{\mcitedefaultseppunct}\relax
\EndOfBibitem
\bibitem[Chen \latin{et~al.}(2018)Chen, He, Ali, Lee, Fong, and
  Law]{chen_prb_2018}
Chen,~C.-Z.; He,~J.~J.; Ali,~M.~N.; Lee,~G.-H.; Fong,~K.~C.; Law,~K.~T.
  Asymmetric Josephson effect in inversion symmetry breaking topological
  materials. \emph{Phys. Rev. B} \textbf{2018}, \emph{075430}, 98\relax
\mciteBstWouldAddEndPuncttrue
\mciteSetBstMidEndSepPunct{\mcitedefaultmidpunct}
{\mcitedefaultendpunct}{\mcitedefaultseppunct}\relax
\EndOfBibitem
\bibitem[Turini \latin{et~al.}(2022)Turini, Salimian, Carrega, Iorio,
  Strambini, Giazotto, Zannier, Sorba, and Heun]{turini_arx_2022}
Turini,~B.; Salimian,~S.; Carrega,~M.; Iorio,~A.; Strambini,~E.; Giazotto,~F.;
  Zannier,~V.; Sorba,~L.; Heun,~S. Josephson diode effect in high mobility InSb
  nanoflags. \emph{arXiv:2207.0877} \textbf{2022}, \relax
\mciteBstWouldAddEndPunctfalse
\mciteSetBstMidEndSepPunct{\mcitedefaultmidpunct}
{}{\mcitedefaultseppunct}\relax
\EndOfBibitem
\bibitem[Wu \latin{et~al.}(2022)Wu, Wang, Xu, Sivakumar, Pasco, Filippozzi,
  Parkin, Zeng, McQueen, and Ali]{wu_nat_2022}
Wu,~H.; Wang,~Y.; Xu,~Y.; Sivakumar,~P.~K.; Pasco,~C.; Filippozzi,~U.;
  Parkin,~S. S.~P.; Zeng,~Y.-J.; McQueen,~T.; Ali,~M.~N. The field-free
  Josephson diode in a van der Waals heterostructure. \emph{Nature}
  \textbf{2022}, \emph{604}, 653--656\relax
\mciteBstWouldAddEndPuncttrue
\mciteSetBstMidEndSepPunct{\mcitedefaultmidpunct}
{\mcitedefaultendpunct}{\mcitedefaultseppunct}\relax
\EndOfBibitem
\bibitem[G{\"u}l \latin{et~al.}(2017)G{\"u}l, Zhang, de~Vries, van Veen, Zuo,
  Mourik, Conesa-Boj, Nowak, van Woerkom, Quintero-P{\'e}rez, Cassidy, Geresdi,
  Koelling, Car, Plissard, Bakkers, and Kouwenhoven]{gul_nanolett_2017}
G{\"u}l,~{\"O}. \latin{et~al.}  Hard superconducting gap in InSb nanowires.
  \emph{Nano Lett.} \textbf{2017}, \emph{17}, 2690--2696\relax
\mciteBstWouldAddEndPuncttrue
\mciteSetBstMidEndSepPunct{\mcitedefaultmidpunct}
{\mcitedefaultendpunct}{\mcitedefaultseppunct}\relax
\EndOfBibitem
\bibitem[Kanne \latin{et~al.}(2021)Kanne, Marnauza, Olsteins, Carrad, Sestoft,
  de~Bruijckere, Zeng, Johnson, Olsson, Grove-Rasmussen, and
  Nyg{\aa}rd]{kanne_natnano_2021}
Kanne,~T.; Marnauza,~M.; Olsteins,~D.; Carrad,~D.~J.; Sestoft,~J.~E.;
  de~Bruijckere,~J.; Zeng,~L.; Johnson,~E.; Olsson,~E.; Grove-Rasmussen,~K.;
  Nyg{\aa}rd,~J. Epitaxial Pb on InAs nanowires for quantum devices. \emph{Nat.
  Nanotechnol.} \textbf{2021}, \emph{16}, 776--781\relax
\mciteBstWouldAddEndPuncttrue
\mciteSetBstMidEndSepPunct{\mcitedefaultmidpunct}
{\mcitedefaultendpunct}{\mcitedefaultseppunct}\relax
\EndOfBibitem
\bibitem[Pendharkar \latin{et~al.}(2021)Pendharkar, Zhang, Wu, Zarassi, Zhang,
  Dempsey, Lee, Harrington, Badawy, Gazibegovic, Jung, Chen, Verheijen,
  Hocevar, Bakkers, Palmstr{\o}m, and Frolov]{pendharkar_science_2021}
Pendharkar,~M. \latin{et~al.}  Parity-preserving and magnetic field–resilient
  superconductivity in InSb nanowires with Sn shells. \emph{Science}
  \textbf{2021}, \emph{372}, 508--511\relax
\mciteBstWouldAddEndPuncttrue
\mciteSetBstMidEndSepPunct{\mcitedefaultmidpunct}
{\mcitedefaultendpunct}{\mcitedefaultseppunct}\relax
\EndOfBibitem
\bibitem[Mazur \latin{et~al.}(2022)Mazur, van Loo, Wang, Dvir, Wang, Khindanov,
  Korneychuk, Borsoi, Dekker, Badawy, Vinke, Gazibegovic, Bakkers,
  Quintero-P{\'e}rez, Heedt, and Kouwenhoven]{mazur_advmater_2022}
Mazur,~G.~P. \latin{et~al.}  Spin-mixing enhanced proximity effect in
  aluminum-based superconductor–semiconductor hybrids. \emph{Adv. Mater.}
  \textbf{2022}, \emph{34}, 1--8\relax
\mciteBstWouldAddEndPuncttrue
\mciteSetBstMidEndSepPunct{\mcitedefaultmidpunct}
{\mcitedefaultendpunct}{\mcitedefaultseppunct}\relax
\EndOfBibitem
\bibitem[Heedt \latin{et~al.}(2021)Heedt, Quintero-P{\'e}rez, Borsoi, Fursina,
  van Loo, Mazur, Nowak, Ammerlaan, Li, Korneychuk, Shen, van~de Poll, Badawy,
  Gazibegovic, de~Jong, Aseev, van Hoogdalem, and
  Kouwenhoven]{heedt_natcomm_2021}
Heedt,~S. \latin{et~al.}  Shadow-wall lithography of ballistic
  superconductor–semiconductor quantum devices. \emph{Nat. Commun.}
  \textbf{2021}, \emph{12}, 4914\relax
\mciteBstWouldAddEndPuncttrue
\mciteSetBstMidEndSepPunct{\mcitedefaultmidpunct}
{\mcitedefaultendpunct}{\mcitedefaultseppunct}\relax
\EndOfBibitem
\bibitem[Zuo \latin{et~al.}(2017)Zuo, Mourik, Szombati, Nijholt, van Woerkom,
  Geresdi, Chen, Ostroukh, Akhmerov, Plissard, Car, Bakkers, Pikulin,
  Kouwenhoven, and Frolov]{zuo_prl_2017}
Zuo,~K.; Mourik,~V.; Szombati,~D.~B.; Nijholt,~B.; van Woerkom,~D.~J.;
  Geresdi,~A.; Chen,~J.; Ostroukh,~V.~P.; Akhmerov,~A.~R.; Plissard,~S.~R.;
  Car,~D.; Bakkers,~E. P. A.~M.; Pikulin,~D.~I.; Kouwenhoven,~L.~P.;
  Frolov,~S.~M. Supercurrent interference in few-mode nanowire Josephson
  junctions. \emph{Phys. Rev. Lett.} \textbf{2017}, \emph{119}, 187704\relax
\mciteBstWouldAddEndPuncttrue
\mciteSetBstMidEndSepPunct{\mcitedefaultmidpunct}
{\mcitedefaultendpunct}{\mcitedefaultseppunct}\relax
\EndOfBibitem
\bibitem[Borsoi \latin{et~al.}(2021)Borsoi, Mazur, van Loo, Nowak, Bourdet, Li,
  Korneychuk, Fursina, Wang, Levajac, Memisevic, Badawy, Gazibegovic, van
  Hoogdalem, Bakkers, Kouwenhoven, Heedt, and
  Quintero-P{\'e}rez]{borsoi_afm_2021}
Borsoi,~F. \latin{et~al.}  Single-shot fabrication of
  semiconducting–superconducting nanowire devices. \emph{Adv. Func. Mater.}
  \textbf{2021}, p. 2102388\relax
\mciteBstWouldAddEndPuncttrue
\mciteSetBstMidEndSepPunct{\mcitedefaultmidpunct}
{\mcitedefaultendpunct}{\mcitedefaultseppunct}\relax
\EndOfBibitem
\bibitem[Badawy \latin{et~al.}(2019)Badawy, Gazibegovic, Borsoi, Heedt, Wang,
  Koelling, Verheijen, Kouwenhoven, and Bakkers]{badawy_nanolett_2019}
Badawy,~G.; Gazibegovic,~S.; Borsoi,~F.; Heedt,~S.; Wang,~C.-A.; Koelling,~S.;
  Verheijen,~M.~A.; Kouwenhoven,~L.~P.; Bakkers,~E. P. A.~M. High mobility
  stemless InSb nanowires. \emph{Nano Lett.} \textbf{2019}, \emph{19},
  3575--3582\relax
\mciteBstWouldAddEndPuncttrue
\mciteSetBstMidEndSepPunct{\mcitedefaultmidpunct}
{\mcitedefaultendpunct}{\mcitedefaultseppunct}\relax
\EndOfBibitem
\bibitem[de~Moor \latin{et~al.}(2018)de~Moor, Bommer, Xu, Winkler, Antipov,
  Bargerbos, Wang, van Loo, het Veld, Gazibegovic, Car, Logan, Pendharkar, Lee,
  Bakkers, Palmstr{\o}m, Lutchyn, Kouwenhoven, and Zhang]{demoor_njphys_2018}
de~Moor,~M. W.~A. \latin{et~al.}  Electric field tunable
  superconductor-semiconductor coupling in Majorana nanowires. \emph{New J.
  Phys.} \textbf{2018}, \emph{20}, 103049\relax
\mciteBstWouldAddEndPuncttrue
\mciteSetBstMidEndSepPunct{\mcitedefaultmidpunct}
{\mcitedefaultendpunct}{\mcitedefaultseppunct}\relax
\EndOfBibitem
\bibitem[Shen \latin{et~al.}(2021)Shen, Winkler, Borsoi, Heedt, Levajac, Wang,
  van Driel, Bouman, Gazibegovic, Veld, Car, Logan, Pendharkar, Palmstr{\o}m,
  Bakkers, Kouwenhoven, and van Heck]{shen_prb_2021}
Shen,~J. \latin{et~al.}  Full parity phase diagram of a proximitized nanowire
  island. \emph{Phys. Rev. B} \textbf{2021}, \emph{104}, 045422\relax
\mciteBstWouldAddEndPuncttrue
\mciteSetBstMidEndSepPunct{\mcitedefaultmidpunct}
{\mcitedefaultendpunct}{\mcitedefaultseppunct}\relax
\EndOfBibitem
\bibitem[Mikkelsen \latin{et~al.}(2018)Mikkelsen, Kotetes, Krogstrup, and
  Flensberg]{mikkelsen_prx_2018}
Mikkelsen,~A.~E.; Kotetes,~P.; Krogstrup,~P.; Flensberg,~K. Hybridization at
  superconductor-semiconductor interfaces. \emph{Phys. Rev. X} \textbf{2018},
  \emph{8}, 031040\relax
\mciteBstWouldAddEndPuncttrue
\mciteSetBstMidEndSepPunct{\mcitedefaultmidpunct}
{\mcitedefaultendpunct}{\mcitedefaultseppunct}\relax
\EndOfBibitem
\bibitem[Antipov \latin{et~al.}(2018)Antipov, Bargerbos, Winkler, Bauer, Rossi,
  and Lutchyn]{antipov_prx_2018}
Antipov,~A.~E.; Bargerbos,~A.; Winkler,~G.~W.; Bauer,~B.; Rossi,~E.;
  Lutchyn,~R.~M. Effects of gate-induced electric fields on semiconductor
  Majorana nanowires. \emph{Phys. Rev. X} \textbf{2018}, \emph{8}, 031041\relax
\mciteBstWouldAddEndPuncttrue
\mciteSetBstMidEndSepPunct{\mcitedefaultmidpunct}
{\mcitedefaultendpunct}{\mcitedefaultseppunct}\relax
\EndOfBibitem
\bibitem[Gharavi and Baugh(2015)Gharavi, and Baugh]{gharavi_prb_2015}
Gharavi,~K.; Baugh,~J. Orbital Josephson interference in a nanowire
  proximity-effect junction. \emph{Phys. Rev. B} \textbf{2015}, \emph{91},
  245436\relax
\mciteBstWouldAddEndPuncttrue
\mciteSetBstMidEndSepPunct{\mcitedefaultmidpunct}
{\mcitedefaultendpunct}{\mcitedefaultseppunct}\relax
\EndOfBibitem
\bibitem[Wang \latin{et~al.}(2022)Wang, Schrade, Levajac, van Driel, Li,
  Gazibegovic, Badawy, het Veld, Lee, Pendharkar, Dempsey, Palmstr{\o}m,
  Bakkers, Fu, Kouwenhoven, and Shen]{wang_sciadv_2022}
Wang,~J.-Y. \latin{et~al.}  Supercurrent parity meter in a nanowire Cooper pair
  transistor. \emph{Sci. Adv.} \textbf{2022}, eabm9896\relax
\mciteBstWouldAddEndPuncttrue
\mciteSetBstMidEndSepPunct{\mcitedefaultmidpunct}
{\mcitedefaultendpunct}{\mcitedefaultseppunct}\relax
\EndOfBibitem
\bibitem[Liu \latin{et~al.}(2021)Liu, van Heck, and Wimmer]{cxliu_prb_2021}
Liu,~C.-X.; van Heck,~B.; Wimmer,~M. Josephson current via an isolated Majorana
  zero mode. \emph{Phys. Rev. B} \textbf{2021}, \emph{103}, 014510\relax
\mciteBstWouldAddEndPuncttrue
\mciteSetBstMidEndSepPunct{\mcitedefaultmidpunct}
{\mcitedefaultendpunct}{\mcitedefaultseppunct}\relax
\EndOfBibitem
\bibitem[Schrade and Fu(2018)Schrade, and Fu]{constantin_prl_2018}
Schrade,~C.; Fu,~L. Parity-controlled $2\pi$ Josephson effect mediated by
  Majorana Kramers pairs. \emph{Phys. Rev. Lett.} \textbf{2018}, \emph{120},
  267002\relax
\mciteBstWouldAddEndPuncttrue
\mciteSetBstMidEndSepPunct{\mcitedefaultmidpunct}
{\mcitedefaultendpunct}{\mcitedefaultseppunct}\relax
\EndOfBibitem
\end{mcitethebibliography}


\providecommand{\latin}[1]{#1}
\makeatletter
\providecommand{\doi}
  {\begingroup\let\do\@makeother\dospecials
  \catcode`\{=1 \catcode`\}=2 \doi@aux}
\providecommand{\doi@aux}[1]{\endgroup\texttt{#1}}
\makeatother
\providecommand*\mcitethebibliography{\thebibliography}
\csname @ifundefined\endcsname{endmcitethebibliography}
  {\let\endmcitethebibliography\endthebibliography}{}
\begin{mcitethebibliography}{7}
\providecommand*\natexlab[1]{#1}
\providecommand*\mciteSetBstSublistMode[1]{}
\providecommand*\mciteSetBstMaxWidthForm[2]{}
\providecommand*\mciteBstWouldAddEndPuncttrue
  {\def\EndOfBibitem{\unskip.}}
\providecommand*\mciteBstWouldAddEndPunctfalse
  {\let\EndOfBibitem\relax}
\providecommand*\mciteSetBstMidEndSepPunct[3]{}
\providecommand*\mciteSetBstSublistLabelBeginEnd[3]{}
\providecommand*\EndOfBibitem{}
\mciteSetBstSublistMode{f}
\mciteSetBstMaxWidthForm{subitem}{(\alph{mcitesubitemcount})}
\mciteSetBstSublistLabelBeginEnd
  {\mcitemaxwidthsubitemform\space}
  {\relax}
  {\relax}

\bibitem[Badawy \latin{et~al.}(2019)Badawy, Gazibegovic, Borsoi, Heedt, Wang,
  Koelling, Verheijen, Kouwenhoven, and Bakkers]{badawy_nanolett_2019}
Badawy,~G.; Gazibegovic,~S.; Borsoi,~F.; Heedt,~S.; Wang,~C.-A.; Koelling,~S.;
  Verheijen,~M.~A.; Kouwenhoven,~L.~P.; Bakkers,~E. P. A.~M. High mobility
  stemless InSb nanowires. \emph{Nano Lett.} \textbf{2019}, \emph{19},
  3575--3582\relax
\mciteBstWouldAddEndPuncttrue
\mciteSetBstMidEndSepPunct{\mcitedefaultmidpunct}
{\mcitedefaultendpunct}{\mcitedefaultseppunct}\relax
\EndOfBibitem
\bibitem[Heedt \latin{et~al.}(2021)Heedt, Quintero-P{\'e}rez, Borsoi, Fursina,
  van Loo, Mazur, Nowak, Ammerlaan, Li, Korneychuk, Shen, van~de Poll, Badawy,
  Gazibegovic, de~Jong, Aseev, van Hoogdalem, and
  Kouwenhoven]{heedt_natcomm_2021}
Heedt,~S. \latin{et~al.}  Shadow-wall lithography of ballistic
  superconductor–semiconductor quantum devices. \emph{Nat. Commun.}
  \textbf{2021}, \emph{12}, 4914\relax
\mciteBstWouldAddEndPuncttrue
\mciteSetBstMidEndSepPunct{\mcitedefaultmidpunct}
{\mcitedefaultendpunct}{\mcitedefaultseppunct}\relax
\EndOfBibitem
\bibitem[Borsoi \latin{et~al.}(2021)Borsoi, Mazur, van Loo, Nowak, Bourdet, Li,
  Korneychuk, Fursina, Wang, Levajac, Memisevic, Badawy, Gazibegovic, van
  Hoogdalem, Bakkers, Kouwenhoven, Heedt, and
  Quintero-P{\'e}rez]{borsoi_afm_2021}
Borsoi,~F. \latin{et~al.}  Single-shot fabrication of
  semiconducting–superconducting nanowire devices. \emph{Adv. Func. Mater.}
  \textbf{2021}, p. 2102388\relax
\mciteBstWouldAddEndPuncttrue
\mciteSetBstMidEndSepPunct{\mcitedefaultmidpunct}
{\mcitedefaultendpunct}{\mcitedefaultseppunct}\relax
\EndOfBibitem
\bibitem[Mazur \latin{et~al.}(2022)Mazur, van Loo, Wang, Dvir, Wang, Khindanov,
  Korneychuk, Borsoi, Dekker, Badawy, Vinke, Gazibegovic, Bakkers,
  Quintero-P{\'e}rez, Heedt, and Kouwenhoven]{mazur_advmater_2022}
Mazur,~G.~P. \latin{et~al.}  Spin-mixing enhanced proximity effect in
  aluminum-based superconductor–semiconductor hybrids. \emph{Adv. Mater.}
  \textbf{2022}, \emph{34}, 1--8\relax
\mciteBstWouldAddEndPuncttrue
\mciteSetBstMidEndSepPunct{\mcitedefaultmidpunct}
{\mcitedefaultendpunct}{\mcitedefaultseppunct}\relax
\EndOfBibitem
\bibitem[de~Moor \latin{et~al.}(2018)de~Moor, Bommer, Xu, Winkler, Antipov,
  Bargerbos, Wang, van Loo, het Veld, Gazibegovic, Car, Logan, Pendharkar, Lee,
  Bakkers, Palmstr{\o}m, Lutchyn, Kouwenhoven, and Zhang]{demoor_njphys_2018}
de~Moor,~M. W.~A. \latin{et~al.}  Electric field tunable
  superconductor-semiconductor coupling in Majorana nanowires. \emph{New J.
  Phys.} \textbf{2018}, \emph{20}, 103049\relax
\mciteBstWouldAddEndPuncttrue
\mciteSetBstMidEndSepPunct{\mcitedefaultmidpunct}
{\mcitedefaultendpunct}{\mcitedefaultseppunct}\relax
\EndOfBibitem
\bibitem[Shen \latin{et~al.}(2021)Shen, Winkler, Borsoi, Heedt, Levajac, Wang,
  van Driel, Bouman, Gazibegovic, Veld, Car, Logan, Pendharkar, Palmstr{\o}m,
  Bakkers, Kouwenhoven, and van Heck]{shen_prb_2021}
Shen,~J. \latin{et~al.}  Full parity phase diagram of a proximitized nanowire
  island. \emph{Phys. Rev. B} \textbf{2021}, \emph{104}, 045422\relax
\mciteBstWouldAddEndPuncttrue
\mciteSetBstMidEndSepPunct{\mcitedefaultmidpunct}
{\mcitedefaultendpunct}{\mcitedefaultseppunct}\relax
\EndOfBibitem
\end{mcitethebibliography}
\end{document}


\section{Methods}

The study in the main text is based on nine InSb-Al nanowire Josephson junction (JJ) devices (Device 1-9) and one InSb-Al nanowire superconducting quantum interference device (SQUID). The JJ devices are used to investigate the impact of junction length on the supercurrent resilience against magnetic field. The SQUID is used to demonstrate how JJs hosting resilient supercurrent can be embedded into a superconducting loop to yield supercurrent interference at high magnetic field. As an additional measurement, the supercurrent resilience against magnetic field is examined in an additional JJ device (Device 10), that is the single arm of the SQUID.    

\subsection{Device fabrication}
All devices in this work were fabricated on $p^+$-doped Si wafers covered with $\sim 300\,\mathrm{nm}$ of thermal $\mathrm{SiO_2}$. For Device 1-9, the thermal $\mathrm{SiO_2}$ is used as a global back gate dielectric. For the SQUID, extra steps in the substrate fabrication were taken in order to create local bottom gates. On top of the thermal $\mathrm{SiO_2}$, the local bottom gates were lithographically defined and produced by depositing $3/17\,\mathrm{nm}$ of Ti/Pd by electron beam evaporation.  Then, $\sim 20\,\mathrm{nm}$ of high-quality $\mathrm{HfO_2}$ layer was grown by atomic layer deposition (ALD) at $110\,\mathrm{\degree C}$ to act as the bottom gate dielectric.

Dielectric structures corresponding to specific shadow-wall patterns were defined by electron-beam lithography on top of the thermal $\mathrm{SiO_2}$ and ALD $\mathrm{HfO_2}$ for the Device 1-9 and the SQUID, respectively. Namely, FOx-25 (HSQ) was spun at $1.5\,\mathrm{krpm}$ for one minute, followed by $2$ minutes of hot baking at $180\,\mathrm{\degree C}$ and patterning lithographically. The HSQ is then developed with MF-321 at $60\,\mathrm{\degree C}$ for $5$ minutes and the substrates are subsequently dried using critical point dryer. This step was followed by the nanowire deposition by an optical nanomanipulator setup and the stemless InSb nanowires \cite{badawy_nanolett_2019} were precisely placed on top of the global back gate (Device 1-9) or the array of local bottom gates (SQUID), close to the HSQ structures. 

Deposition of the superconducting Al film was carried out in the nominally identical steps for all devices in this study. After gentle hydrogen cleaning of the nanowire surface, the superconducting film was grown by directional evaporation of Al. The Al flux in the deposition was $17\,\mathrm{nm}$ and the angle with respect to the substrate was $30\degree$ \citep{heedt_natcomm_2021,borsoi_afm_2021}. Due to the specific angle and the regular hexagonal nanowire cross-section, the Al film continuously covers three nanowire facets, as shown in the above cited references. On one facet the Al film is deposited perpendicularly and the film thickness on this facet is $\sim15\,\mathrm{nm}$, as $\sim2\,\mathrm{nm}$ of Al self-terminately oxidizes in the air. The direction of the Al deposition forms an angle of $30\degree$ with the other two facets and these two facets therefore receive $\sin{30\degree}=0.5$ of the Al flux and have the film thickness of $\sim7\,\mathrm{nm}$ after the oxidation. Lithographically patterned dielectric structures cast shadows during the Al deposition and therefore selectively define the sections along the nanowire where the superconducting film is grown and where the semiconducting junction is formed. Additionally, the arrangement of the shadow-wall structures on the SQUID substrate determines a shadowed substrate area without Al enclosed by the two JJs that represents the superconducting loop of the SQUID. Finally, in all devices the superconducting film on the nanowire facets forms a continuous connection to the substrate and extends to pre-patterned bonding pads such that additional fabrication steps to contact the nanowires are not needed.

In this work, seven nanowire JJ devices (Device 1-7) were fabricated on a single chip, while the other two (Device 8-9) come from other two chips that passed through the nominally identical fabrication steps. The SQUID was fabricated on a separate chip in the fabrication steps as explained above. 

\subsection{Measurement setup}
We perform the electrical transport measurements at $\sim20\,\mathrm{mK}$ base temperature in a dilution refrigerator equipped with a vector-rotate magnet. Source and drain leads of the device are bonded each to two printed circuit board (PCB) pads that are via low-pass filters connected to the fridge lines. In this way each device occupies in total four fridge lines - allowing for measurements in a two- and four-terminal configuration. 

We perform the conductance measurements in the two-terminal voltage bias setup in the standard lock-in configuration. Source and drain are connected to the measurement setup by two fridge lines, while the remaining two fridge lines are kept floated. The voltage bias $V_b$ is swept by a dc-voltage source while the ac-voltage $dV_b=10\mu V$ is set by a lock-in amplifier. The total current $I+dI$ through the sample is measured by a current-meter amplifier. The dc- and the ac-voltage drops over the sample are obtained by subtracting the voltage drops over the series resistance $R_s=8.89k\Omega$ as $V=V_b-IR_s$ and $dV=dV_b-dIR_s$. This series resistance accounts for other resistive elements in the circuit such as the two fridge lines, the resistance of the voltage source and the current-meter amplifier and the resistance of the low-pass filters on the printed circuit board.    
For collecting the data from which the switching current is extracted, four-terminal current-bias setup is used. Two fridge lines are used to connect a current source and apply the dc-current bias $I_b$ through a device, while the other two fridge lines are used to connect a voltage-meter and measure the dc-voltage drop $V$ over the device. The current bias is swept in steps of $20\,\mathrm{pA}$ - $60\,\mathrm{pA}$, depending on the range of current-bias that is applied. As the voltage-meter measures at the room temperature the sum of the voltage drops over the device and the two fridge lines, a dc-offset of $\sim 0.01\,\mathrm{mV}$ are substracted to  compensate for the difference in the thermal voltage drops over the fridge lines.\\

\subsection{Data analysis}

All the codes used for the data analysis in this work are available in the data repository. The details of the data analysis procedures performed in these codes are described in the following subsections.

\subsubsection{Extracting normal state conductance $G_n$}
Normal state conductance $G_n$ is extracted from the data collected in the voltage-bias measurements of the nanowire JJ devices. After correcting for the series resistance $R_s$ (as explained in the previous section), the normal state conductance is obtained as $G_n(V_g)=(G_n^+(V_g)+G_n^-(V_g))/2$ where $G_n^+(V_g)=\langle\frac{dI}{dV}(V_g,1\,\mathrm{mV}<V<2\,\mathrm{mV})\rangle$ and $G_n^-(V_g)=\langle\frac{dI}{dV}(V_g,-2\,\mathrm{mV}<V<-1\,\mathrm{mV})\rangle$ are averaged conductances at the positive and the negative source-drain voltages much larger than the double value of the superconducting gap ($2\Delta \sim 500\,\mathrm{\mu V}$). 

\begin{figure}[h] 
\centering
\includegraphics[width=\linewidth]{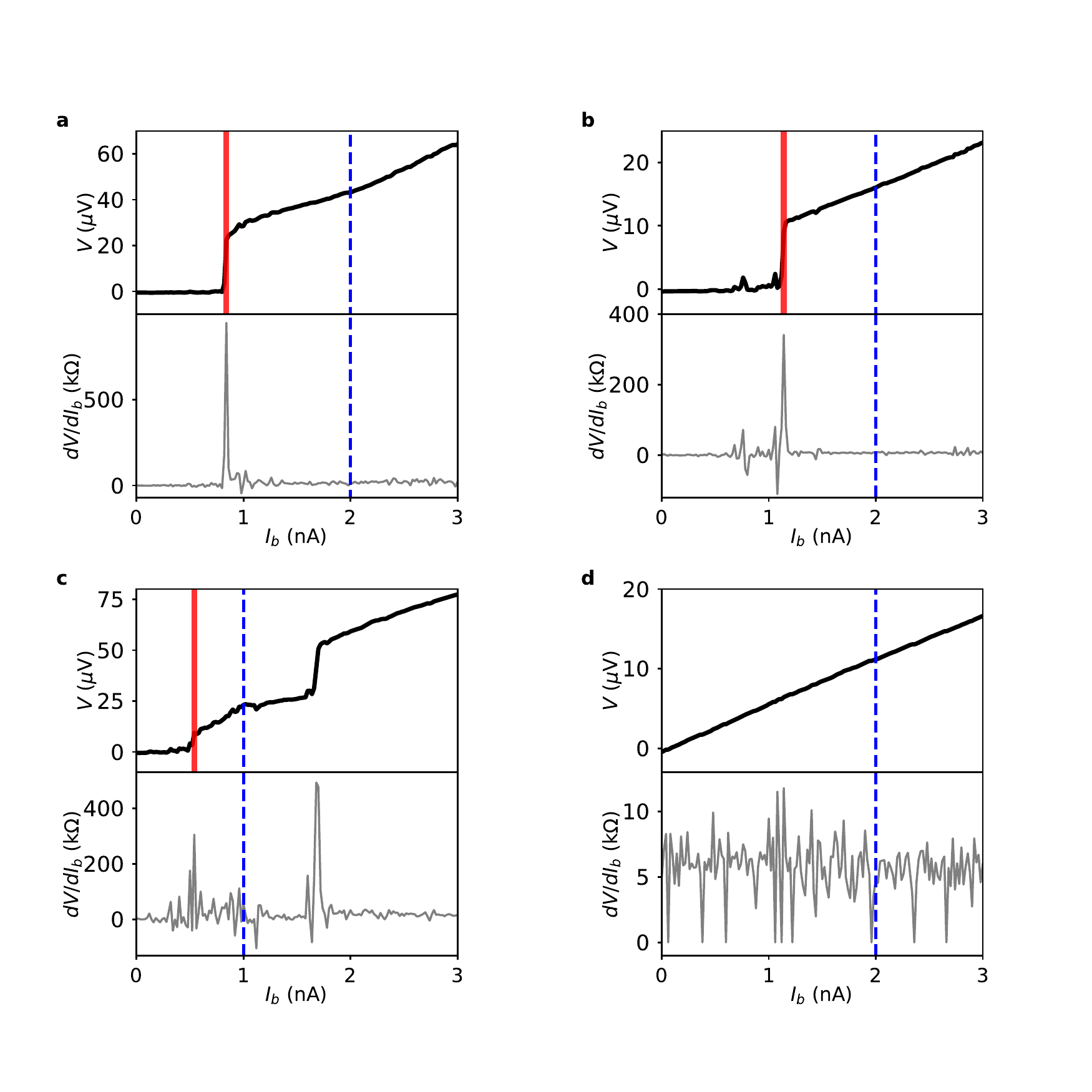} 
\caption{\textbf{Extraction of switching current:} Examples in \textbf{(a)-(d)} show the voltage drop (top) and the numerically calculated differential resistance (bottom) traces as functions of the current bias $I_b$. The extracted switching current $I_{sw}$ (red) and the ranges over which the presence of a switch is examined (blue) are marked by the vertical lines. These traces were taken in Device 2 ($L=31\,\mathrm{nm}$) at $B=1\,\mathrm{T}$ parallel magnetic field.}\label{fig:SIsw}
\end{figure}

\subsubsection{Extracting switching current $I_{sw}$}

Switching current is extracted for each $(V,I_b)$ trace measured in the current-bias setup. Four-examples of $(V,I_b)$ traces are shown in the top parts of Fig. \ref{fig:SIsw}a-d. The corresponding differential resistance $(dV/dI_b,I_b)$ traces are calculated as numerical derivatives and are plotted in the bottom parts of Fig. \ref{fig:SIsw}a-d. The data in Fig. \ref{fig:SIsw} corresponds to four traces from the back gate sweep at parallel magnetic field of $1\,\mathrm{T}$ in Device 2. These traces are chosen to motivate the particular method used in the switching current extraction.     

From a perfectly clean $(V,I_b)$ trace, as the one in Fig. \ref{fig:SIsw}a, with a single voltage step corresponding to the switching current $I_{sw}$, $I_{sw}$ can in principle be extracted by setting a threshold voltage $V_{th}$, such that $V_{th}=V(I_b=I_{sw})$. However, this can give underestimated extracted values as the voltage $V$ can due to noise fluctuate for current bias values lower than the switching current - as shown in the $(V,I_b)$ trace in Fig. \ref{fig:SIsw}b. Setting higher $V_{th}$ values to prevent this, can, on the other hand, give an overestimation of the extracted value if the switching current is small. Therefore, when extracting $I_{sw}$, we rather look at the maximum in the differential resistance, as it resembles the sharpness of a switch in a $(V,I_b)$ trace.

For each differential resistance $(dV/dI_b,I_b)$ trace, the maximal value (peak) of $dV/dI_b$ is found and divided by the third value of the same $(dV/dI_b,I_b)$ trace sorted in decreasing order. In this way we quantify how dominant the peak in the differential resistance is. If the obtained value is smaller than the analogous value obtained from the trace in Fig. \ref{fig:SIsw}d with clearly no switch in it - the peak in differential resistance is not dominant and the switching current is extracted as a "not a number" (NaN) value. These NaN values correspond to the interruptions in the red $I_{sw}$ traces plotted over 2D maps throughout the study.

The trace in Fig. \ref{fig:SIsw}c depicts that the range over which the dominant peak in $(dV/dI_b,I_b)$ is searched for can affect the extracted value. For example, there is a dominant peak in $(dV/dI_b,I_b)$ in Fig. \ref{fig:SIsw}c at $I_b\sim1.7\,\mathrm{nA}$, but it does not correspond to the switching current. Therefore, the range in which the switching current is searched for is an important input parameter that is marked by the blue lines in Fig. \ref{fig:SIsw}. This parameter is commonly set at sufficiently high values and subsequently adjusted for particular traces where it leads to mistakes as the one described in Fig. \ref{fig:SIsw}c. The red lines in Fig. \ref{fig:SIsw} mark the extracted switching current values and nicely match the dominant peaks of the differential resistance in the relevant ranges of the current bias.

The described algorithm successfully identifies the switching current in most of the traces. After applying it, additional corrections were made after checking how an extracted $I_{sw}$ value matches to its corresponding $(V,I_b)$ trace. Some extracted finite $I_{sw}$ values were set then to NaN if found to have been extracted in a highly smeared $(V,I_b)$ trace.  On the other hand, in some non-smeared $(V,I_b)$ traces with NaN extracted $I_{sw}$ values, the switching current is re-extracted by extracting the position of the global maximum in the differential resistance trace. Such post-extraction corrections were performed equally frequently for all devices (5-10\% of all $(V,I_b)$ traces).  

\subsubsection{Extracting critical magnetic field $B_{Ic}$}
By applying the above described algorithm to extract the switching current $I_{sw}$, we extract $I_{sw}(B)$ from the 2D maps shown in Fig. \ref{fig:S6} where the voltage drop $V$ is measured as the current bias $I_b$ and the parallel magnetic field $B$ are swept. By analyzing the evolution of the $(V,I_b)$ linecuts in $B$ field, it can be noticed that the algorithm may give an isolated NaN value for $I_{sw}$ at some $B$ value even if the switching current is correctly extracted at higher fields. Therefore, defining the critical field of switching current $B_{Ic}$ as the lowest $B$ field for which the algorithm gives NaN value for $I_{sw}$ can lead to underestimations of $B_{Ic}$. However, if the algorithm gives NaN values for two consecutive $B$ field values, then even occasionally extracted $I_{sw}$ values different from NaN at higher fields are most often false-positive extracted values. We therefore determine the critical field $B_{Ic}$ as the lowest field such that two consecutive extracted values for $I_{sw}$ are NaN. In Fig. \ref{fig:S6} $I_{sw}$ is plotted up to the determined $B_{Ic}$ while the entire $I_{sw}(B)$ data is available in the data repository.  


\begin{figure}[h] 
\centering
\includegraphics[width=\linewidth] {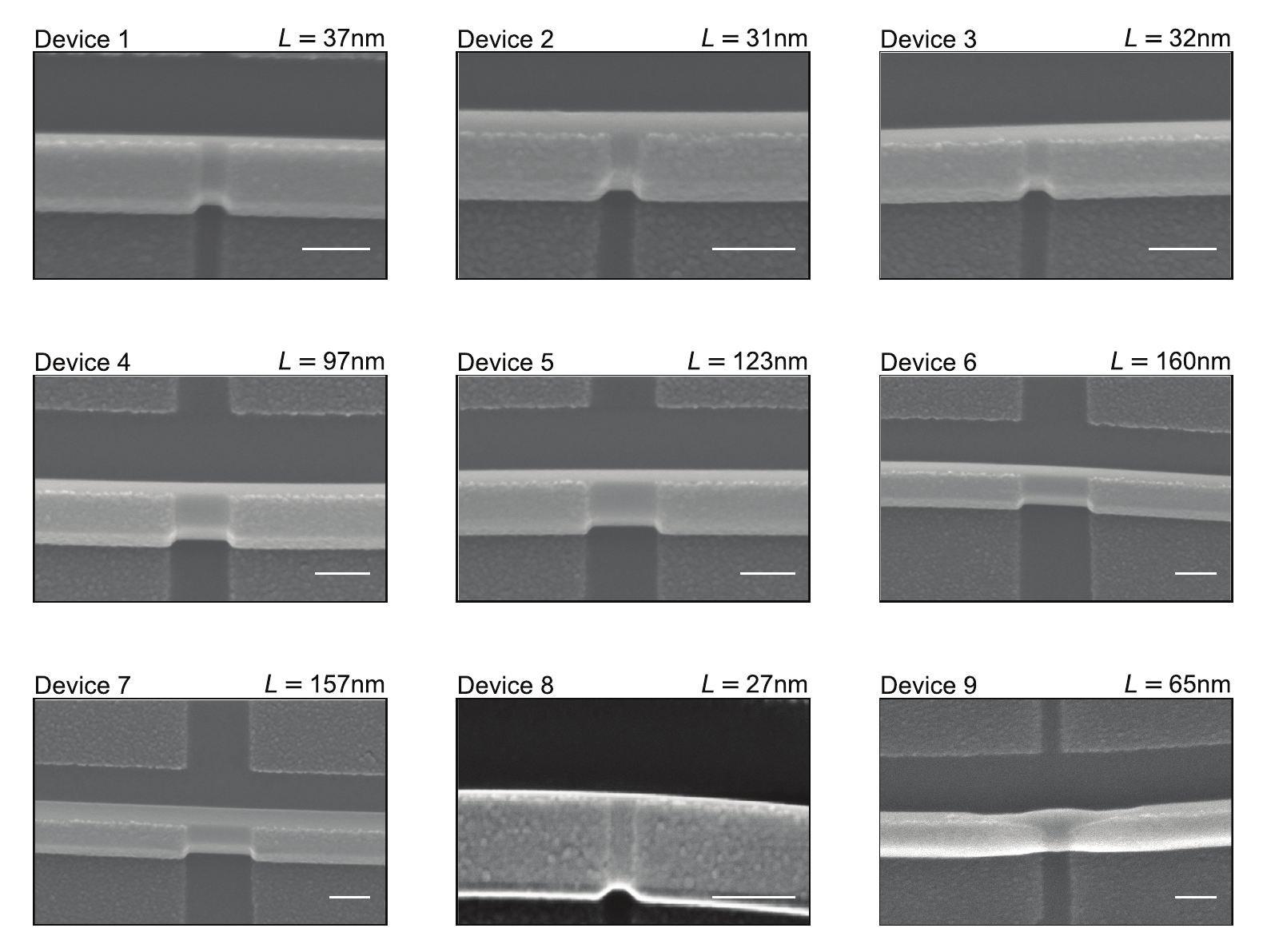}
\caption{\textbf{Nine nanowire Josephson junction devices:} SEM images of the junctions with the corresponding device name (Device 1-9) and the junction length $L$. The diameter of the nanowires is $\sim 100\,\mathrm{nm}$ (between $90\,\mathrm{nm}$ and $110\,\mathrm{nm}$). The scale bars correspond to $100\,\mathrm{nm}$.   
}\label{fig:S1}
\end{figure}

\begin{figure}[h] 
\centering
\includegraphics[width=\linewidth]{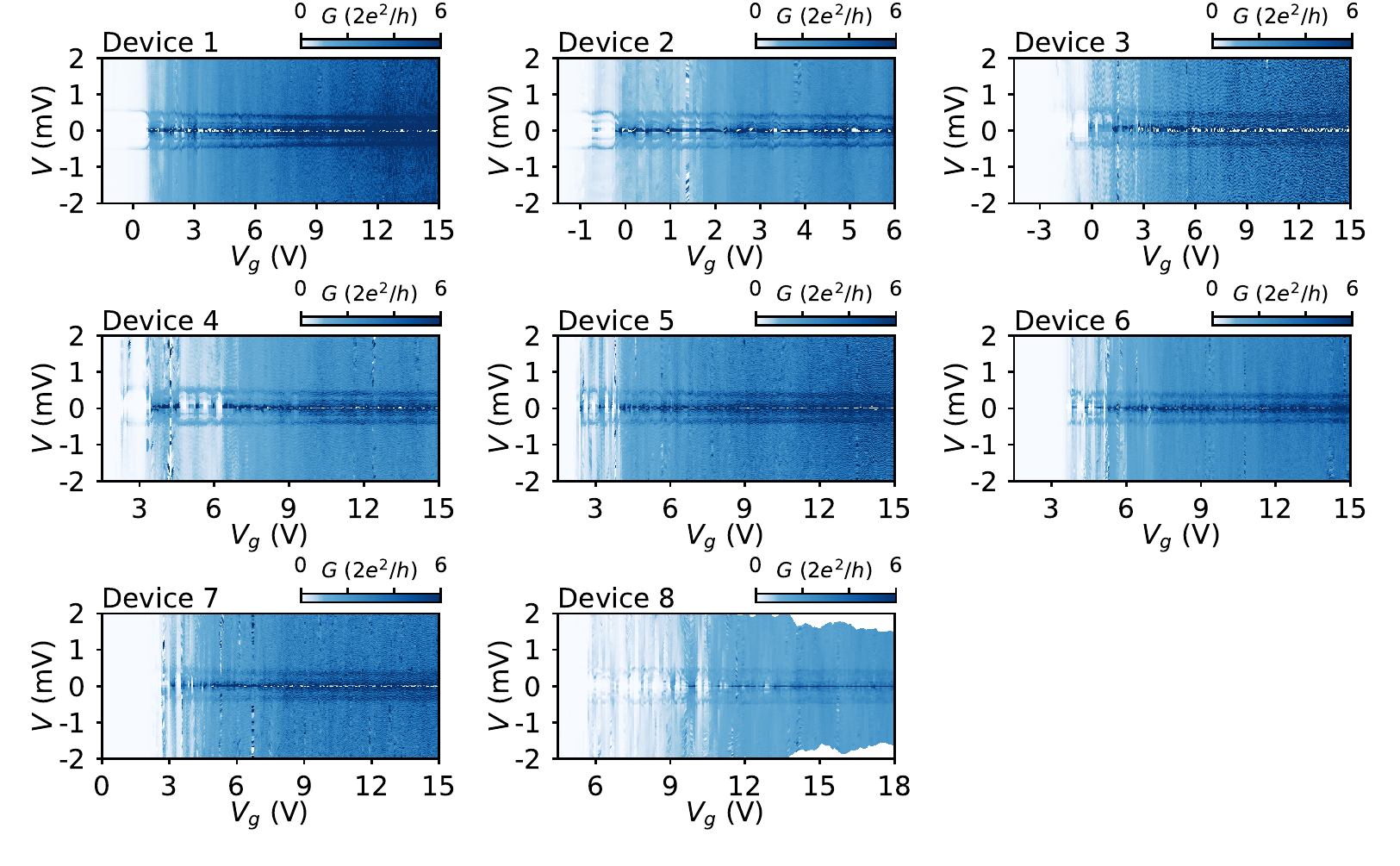}
\caption{\textbf{Differential conductance at zero-field:} Measured differential conductance $G$ through Device 1-8 as a function of the voltage drop $V$ between the source and drain and the back gate voltage $V_g$. The normal state conductance dependences $G_n(V_g)$ for Device 1-8 are obtained from these 2D maps. as described in the Data analysis section. The analogous 2D map was not taken for Device 9 and the $G_n(V_g)$ dependence for this device was measured as a single trace at $V>1\,\mathrm{mV}$. }\label{fig:SVbias}
\end{figure}

\begin{figure}[b] 
\centering
\includegraphics[width=\linewidth] {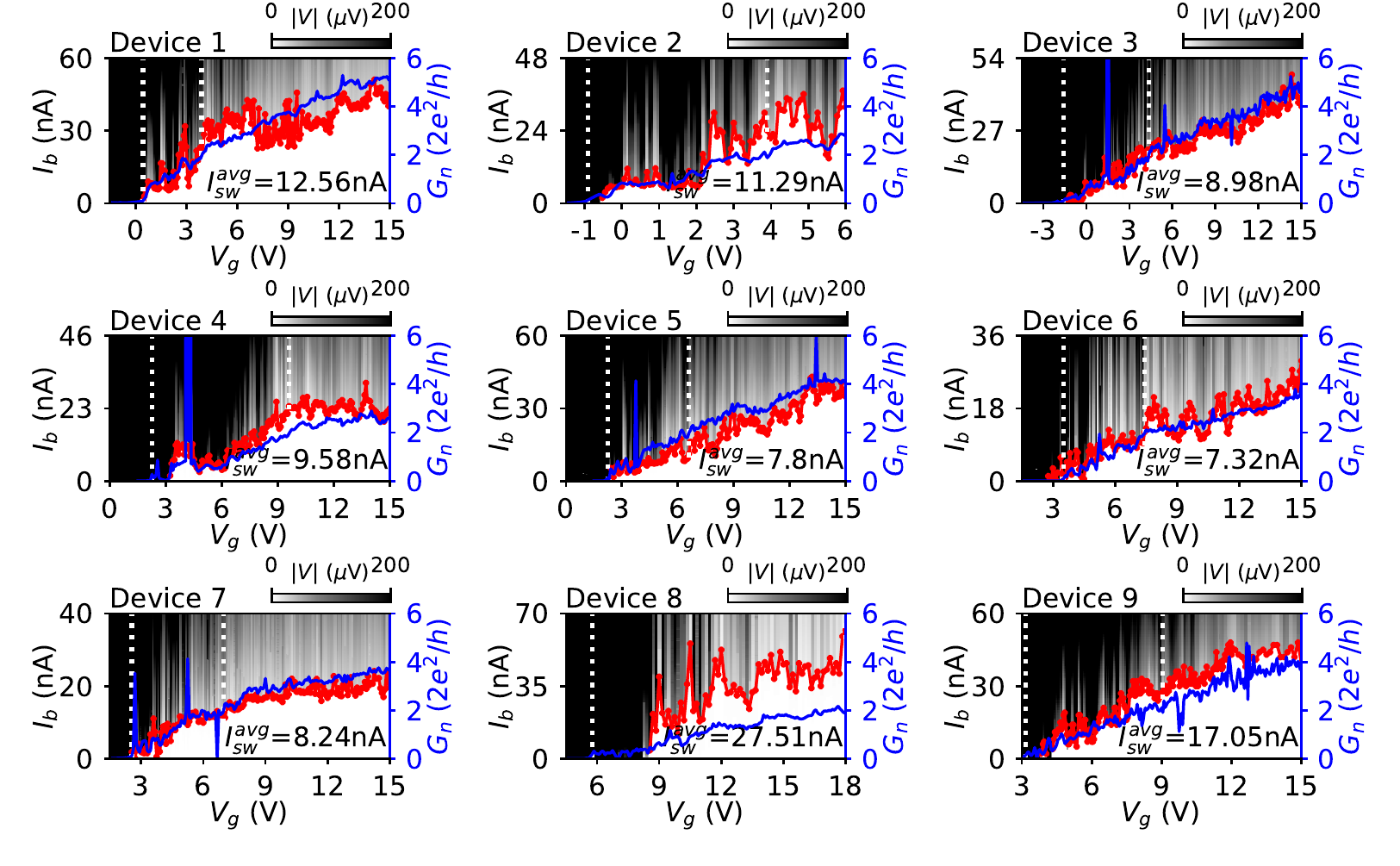}
\caption{\textbf{Tunable switching current and normal conductance at zero-field:} - For Device 1-9 the extracted switching current $I_{sw}$ (red) and measured normal conductance $G_n$ (blue) are plotted over $I_b-V_g$ 2D maps obtained in the current bias measurements at zero-field. All devices show tunability by the back gate voltage $V_g$ from the pinch-off regime with no supercurrent to the open regime with $I_{sw}$ of several tenths of $\mathrm{nA}$ and $G_n$ of few $G_0$ (with $G_0=2e^2/h$). $G_n(V_g)$ dependences are obtained from the data shown in Fig. \ref{fig:SVbias}. The white dotted vertical lines mark the ranges of $V_g$ over which $G_n$ increases from $0.01G_0$  to $2G_0$. The average switching currents in these intervals are shown as insets.}\label{fig:S2}
\end{figure}

\begin{figure}[h] 
\centering
\includegraphics[width=\linewidth] {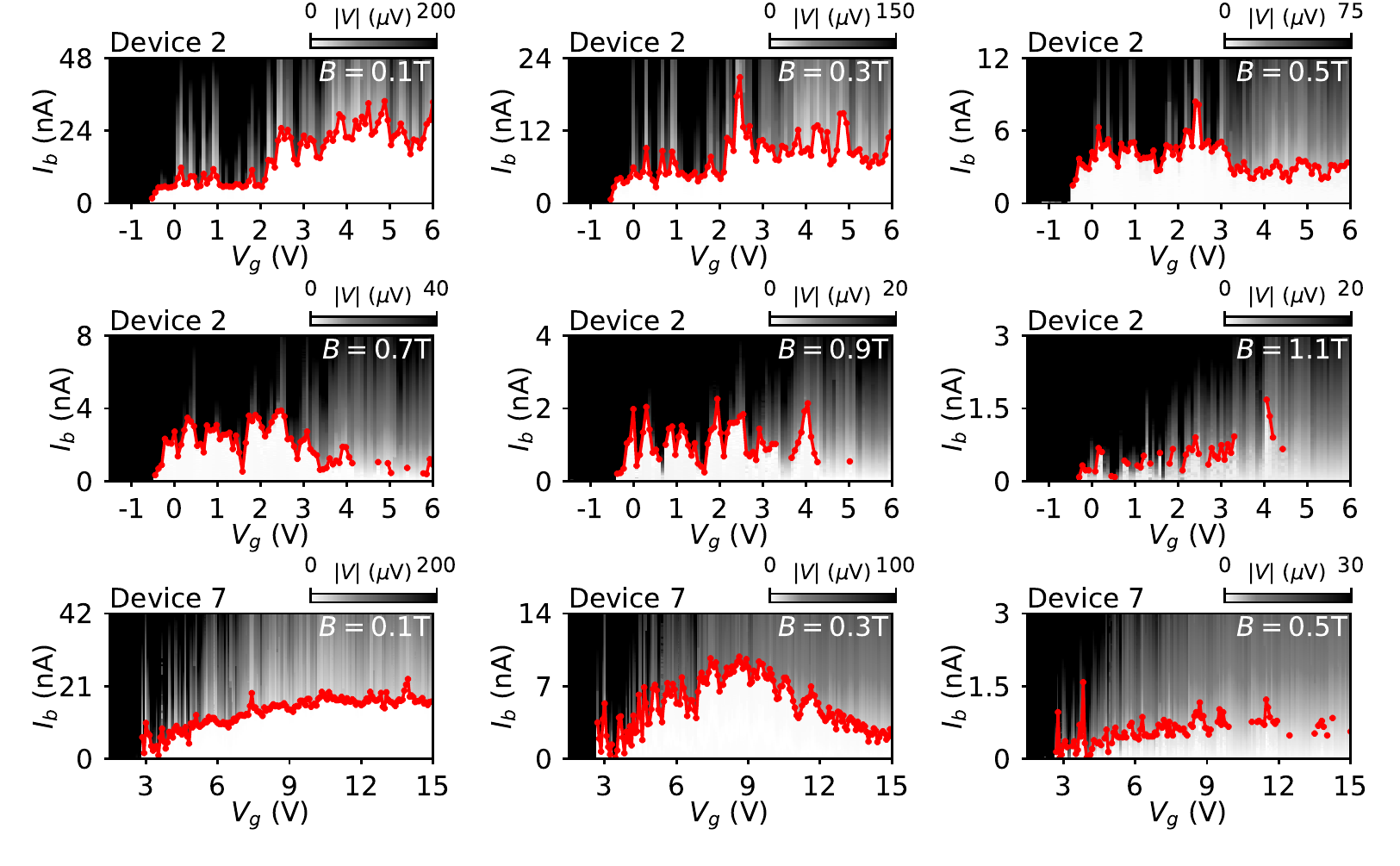}
\caption{\textbf{Background data for Fig. 2:} The extracted switching current $I_{sw}$ (red) as a function of the back gate voltage $V_g$ at several parallel field values. The corresponding parallel magnetic fields are shown as insets.}\label{fig:S34}
\end{figure}

\begin{figure}[h] 
\centering
\includegraphics[width=\linewidth] {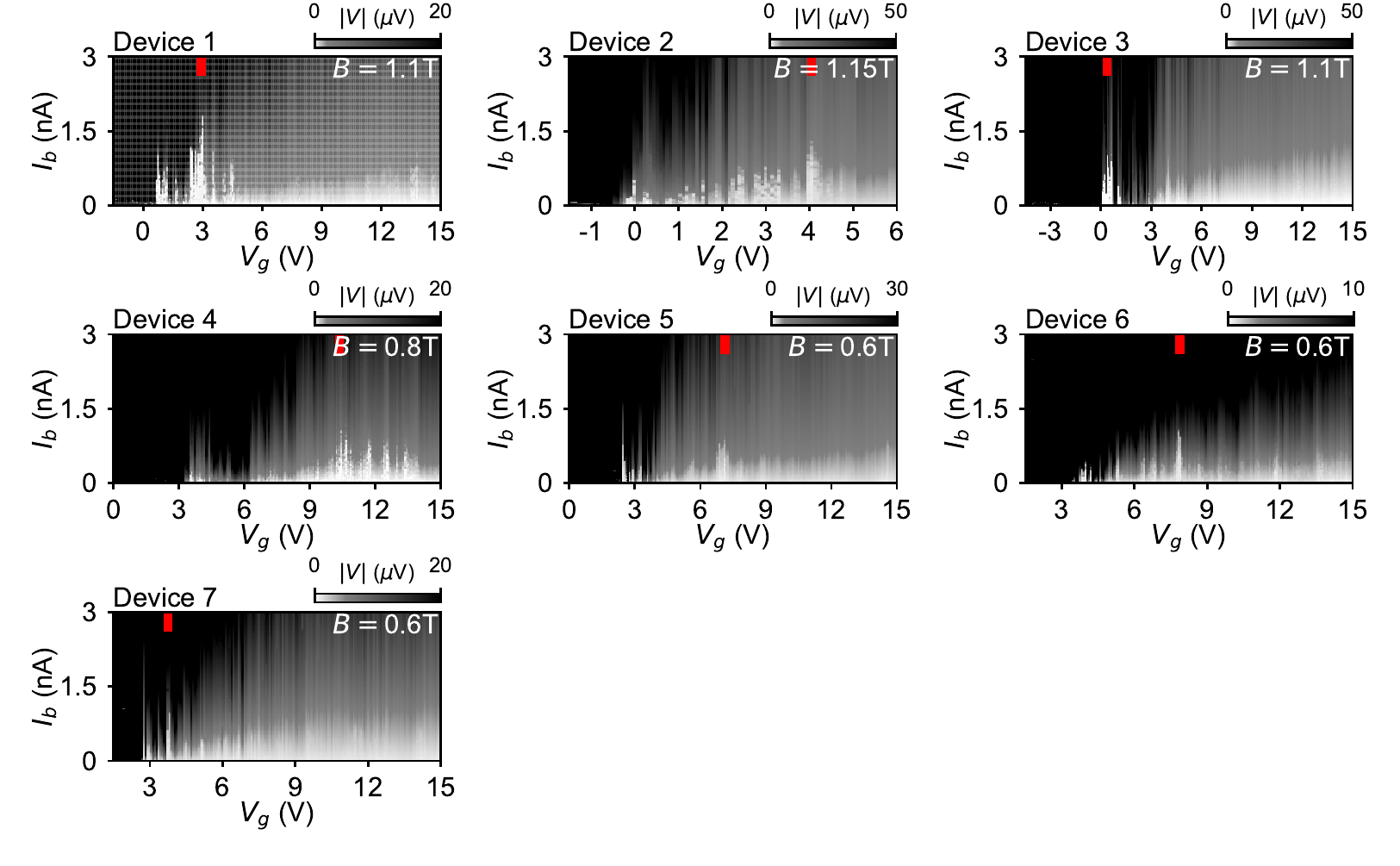}
\caption{\textbf{Identifying the resilient gate settings $V_{g,res}$}: The back gate voltage $V_g$ is swept at high parallel magnetic field for Device 1-7. The red markers denote the resilient gate settings $V_{g,res}$. $V_g$ is set to these values for obtaining the magnetic field dependences shown in Fig. 3 and Fig. \ref{fig:S6}. The analogous measurements were not performed for Device 8-9.}\label{fig:S5}
\end{figure}

\begin{figure}[h] 
\centering
\includegraphics[width=\linewidth] {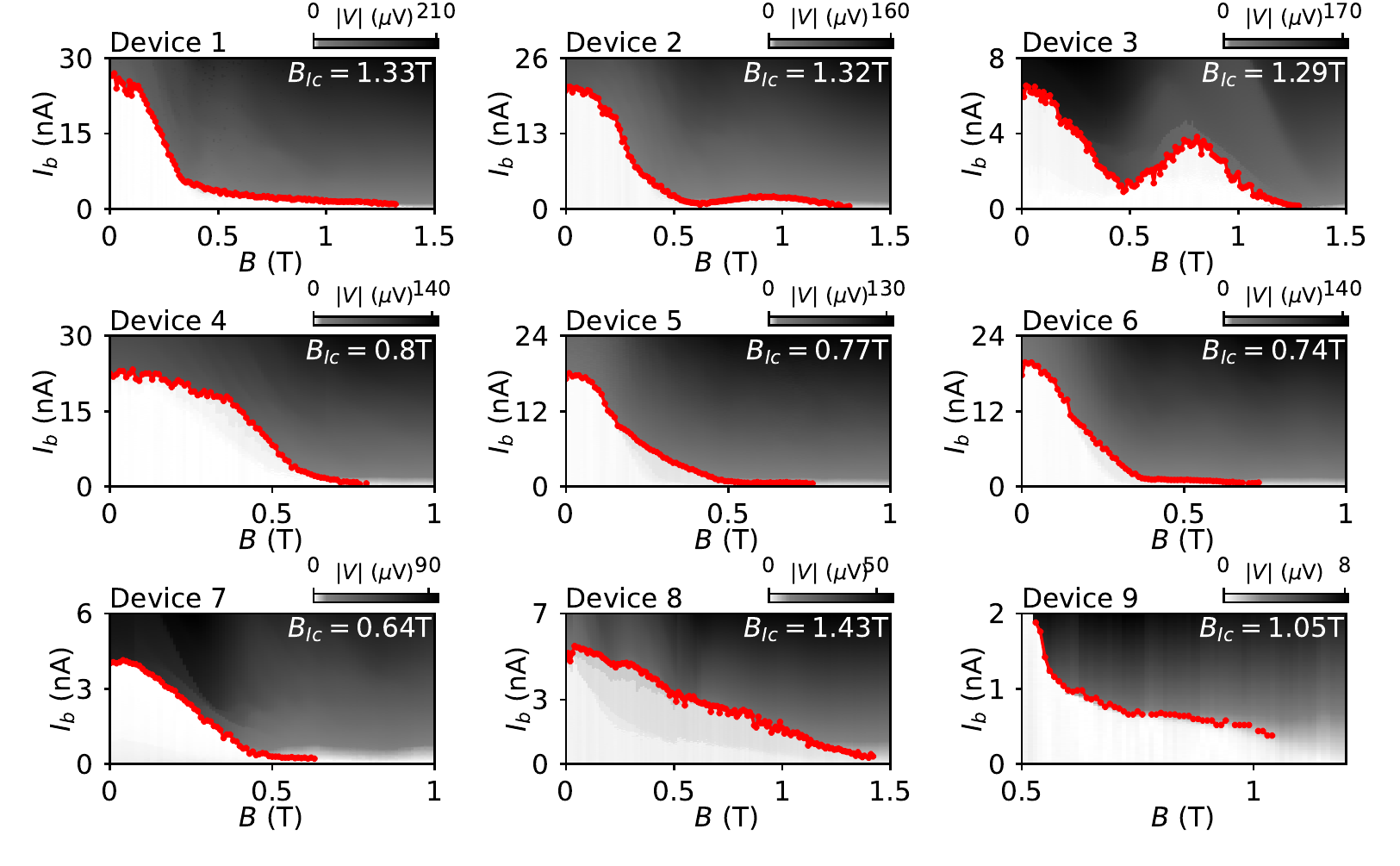}
\caption{\textbf{Evolution of switching current in parallel magnetic field:} Dependence of the switching current $I_{sw}$ (red) on the parallel magnetic field $B$ for Device 1-9. The back gate is set at the resilient gate setting $V_g=V_{g,res}$ for Device 1-7 and at $V_g=15\,\mathrm{V}$ for Device 8-9 (see the Data selection and reproducibility section). The corresponding extracted critical field $B_c$ is shown as an inset. The gate settings for Device 1-7 are marked by the red markers in Fig. \ref{fig:S5}.}\label{fig:S6}
\end{figure}

\begin{figure}[h] 
\centering
\includegraphics[width=\linewidth] {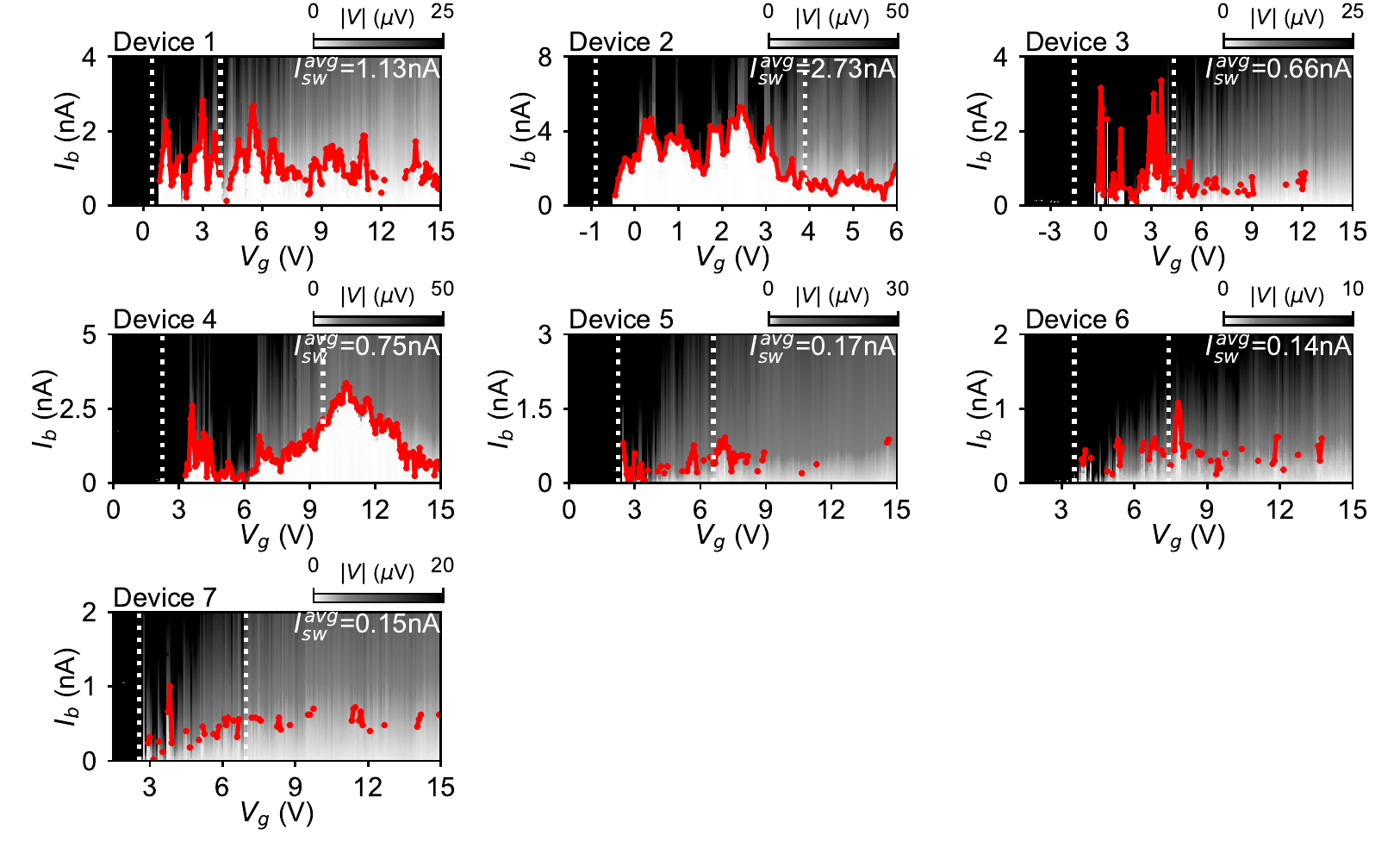}
\caption{\textbf{Switching current at $B=0.6\,\mathrm{T}$ parallel magnetic field:} Dependence of the extracted switching current $I_{sw}$ (red) on the back gate voltage $V_g$ at the parallel field $B=0.6\,\mathrm{T}$ for Device 1-7. The white dotted vertical lines indicate the ranges of $V_g$ over which the normal state conductance $G_n$ at zero-field of the corresponding device increases from $0.01G_0$  to $2G_0$. The average switching currents over these intervals are shown as insets. The analogous measurement was not performed for Device 8-9.}\label{fig:S7}
\end{figure}

\section{Effects of junction length and global back gate on induced superconducting gap}

\begin{figure}[h] 
\centering
\includegraphics[width=\linewidth] {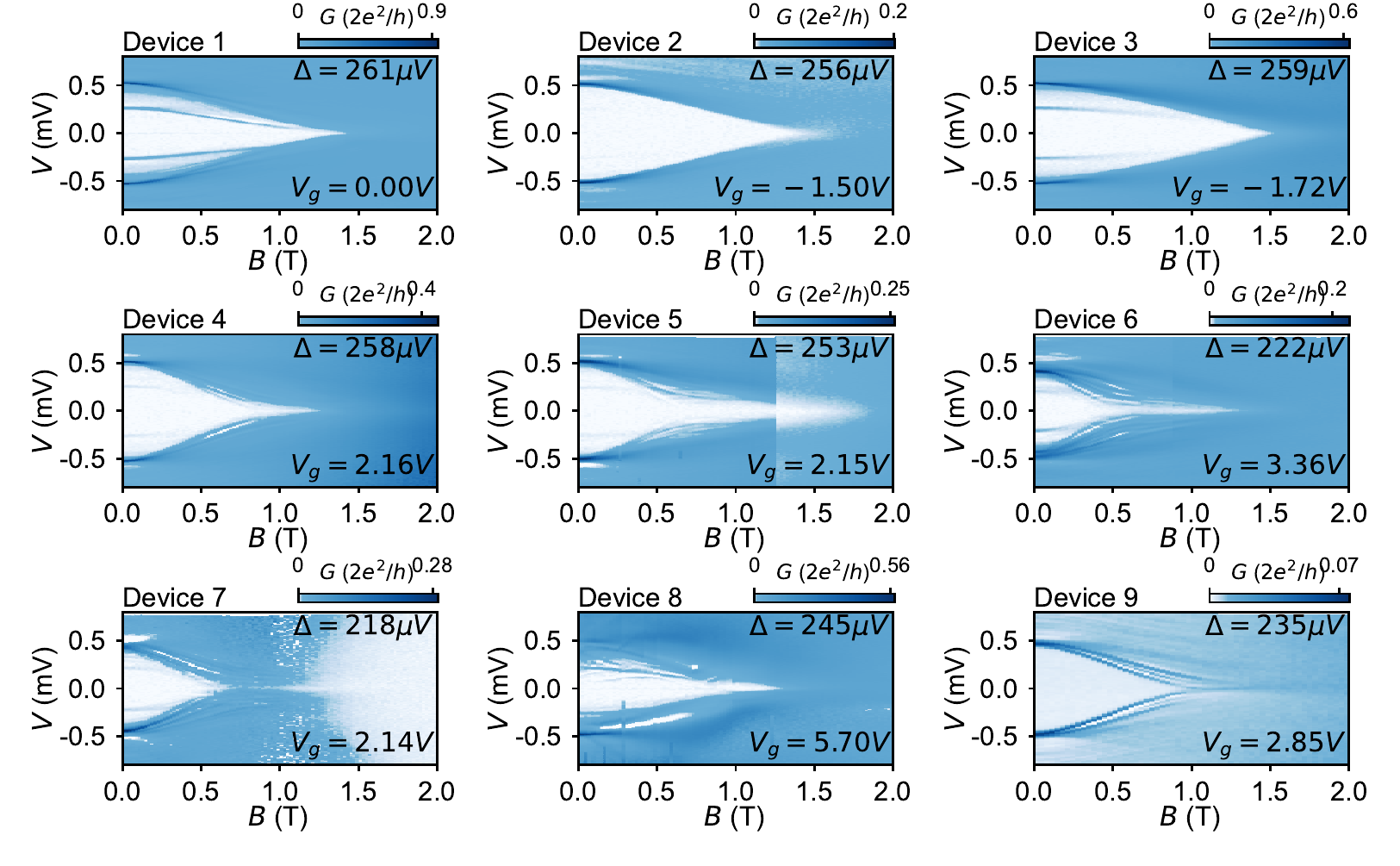}
\caption{\textbf{Evolution of induced superconducting spectra in parallel magnetic field:} Dependence of the tunneling conductance $G$ on the parallel magnetic field $B$ for Device 1-9. Extracted induced superconducting gap at zero field $\Delta$ and the back gate voltage $V_g$ at which the tunneling spectroscopy is measured for each device are shown as insets.}\label{fig:S11}
\end{figure}

In order to measure the induced superconducting gap for Device 1-9 and study its evolution in parallel magnetic field, tunneling spectroscopy is performed in the voltage bias setup. 

In Fig. \ref{fig:S11} the evolution of induced superconducting gap in parallel magnetic field is shown for Device 1-9. Each subfigure represents a 2D map of the tunneling conductance as a function of the voltage drop over the junction and the parallel field. Two coherence peaks corresponding to the double value of the induced gap $\Delta$ appear in the tunneling conductance at $|V|=2\Delta$. By extracting the peak separation and dividing it by 4 for each Device 1-9 at zero field, the values for induced superconducting gap are calculated. These values are shown as insets in Fig. \ref{fig:S11}, together with the global back gate voltage at which the corresponding conductance maps are obtained. 

In Fig. \ref{fig:S11} it can be seen that the three short junctions (Device 1,2 and 3) have larger values of the induced gap with the critical parallel field of $\sim 1.5\,\mathrm{T}$ - similar to the parent superconducting gap in the Al film \cite{borsoi_afm_2021, heedt_natcomm_2021, mazur_advmater_2022}. On the other hand, the two longest junctions (Device 6 and 7) are characterized by reduced induced gaps and subgap states evolving towards zero energy and effectively closing the gap well before the parent superconducting gap vanishes. These differences in the induced gap sizes and their evolution in parallel magnetic field for junctions of different lengths are accompanied by differences in the gate settings at which different devices are set into the tunneling regime. Namely, it can be noticed that shorter devices mostly require low or even negative back gate voltages for reaching the tunneling regime, while this value is higher for the longer junctions. A valid question that arises is whether the differences in the tunneling spectroscopy in Fig. \ref{fig:S11} are due to the differences in the junction lengths or due to the differences in the electrical fields induced by the different gate voltages.

Despite the differences present among the nine devices in the tunneling regime regarding the back gate settings, the junction lengths and the conductance values, some conclusions can be made by looking at specific subsets of the devices for which some of these parameters are comparable. By comparing the data for Device 4, 5 and 7, it can be seen that with almost the same gate settings of $V_g \sim 2.15\,\mathrm{V}$ and the comparable tunneling conductance values $G_n \sim 0.3-0.4\,G_0$, the shortest device out of the three (Device 4) exhibits the largest induced gap that closes at the highest field. The data for the other two devices (Device 5 and 7) suggest that gradual increases of the junction length lead to weaker proximity effect with gradually smaller induced gap and gradually lower critical parallel field of the induced gap. Furthermore, the shortest device in the study (Device 8) requires the largest gate voltage to be tuned into the tunneling regime ($V_g=5.7\,\mathrm{V}$) and still exhibits larger induced gap than the longest devices (Device 6-7) measured at the lower gate voltages. Despite the high gate voltage, the induced gap of Device 8 closes at $\sim 1.3\,\mathrm{T}$. However, in comparison to the remaining short junctions measured at significantly lower gate voltages (Device 1,2 and 3), Device 8 has poorer induced superconducting properties, probably due to the the high gate voltage and reduced superconductor-semiconductor coupling. 

We can conclude that junction length is an important parameter that influences the induced superconducting gap. This does not exclude an effect that the applied back gate voltage has on induced superconductivity. Moreover, the data in Fig. \ref{fig:S11} demonstrates that both the junction length and the back gate voltage determine the semiconductor-superconductor hybridization. This confirms that the electrostatic profile inside a hybrid nanowire JJ device - influenced by both device geometry and gate voltage - can control the strength of the semiconductor-superconductor hybridization \cite{demoor_njphys_2018, shen_prb_2021}{}.  

The stronger proximity effect in the short JJs could originate from an electron layers accumulated at the interfaces between the semiconducting nanowire and the superconducting leads. Namely, the band offset at an InSb-Al interface can cause a bending of the InSb conduction band and results in a strongly proximitized electron layer at the interdace with Al. Because of a finite lateral extension of such layers from the two sides of a short JJ, the junction superconducting properties could be enhanced. Note that in some short JJs in our study the normal conductance and supercurrent have been measured to be finite when no back gate voltage is applied (see the data for Device 2 and 3 in Fig. \ref{fig:S2}). This could suggest that the accumulation layers can fully extend over a $\sim 30\,\mathrm{nm}$ junction by extending $\sim 15\,\mathrm{nm}$ laterally at each side.

The evidence of different strengths of hybridization in junctions of different lengths is in agreement with the reported zero-field values of the induced gap in Fig. \ref{fig:S11} and the average switching current values at zero field in Fig. \ref{fig:S2}. Although the induced gap is characterized in the tunneling regime with no supercurrent, the critical parallel fields of switching current in Fig. 3e in the main text roughly match the parallel field values at which the induced gaps close in Fig. \ref{fig:S11}.

\section{Effects of local gates on supercurrent resilience}

\begin{figure}[h] 
\centering
\includegraphics[width=\linewidth] {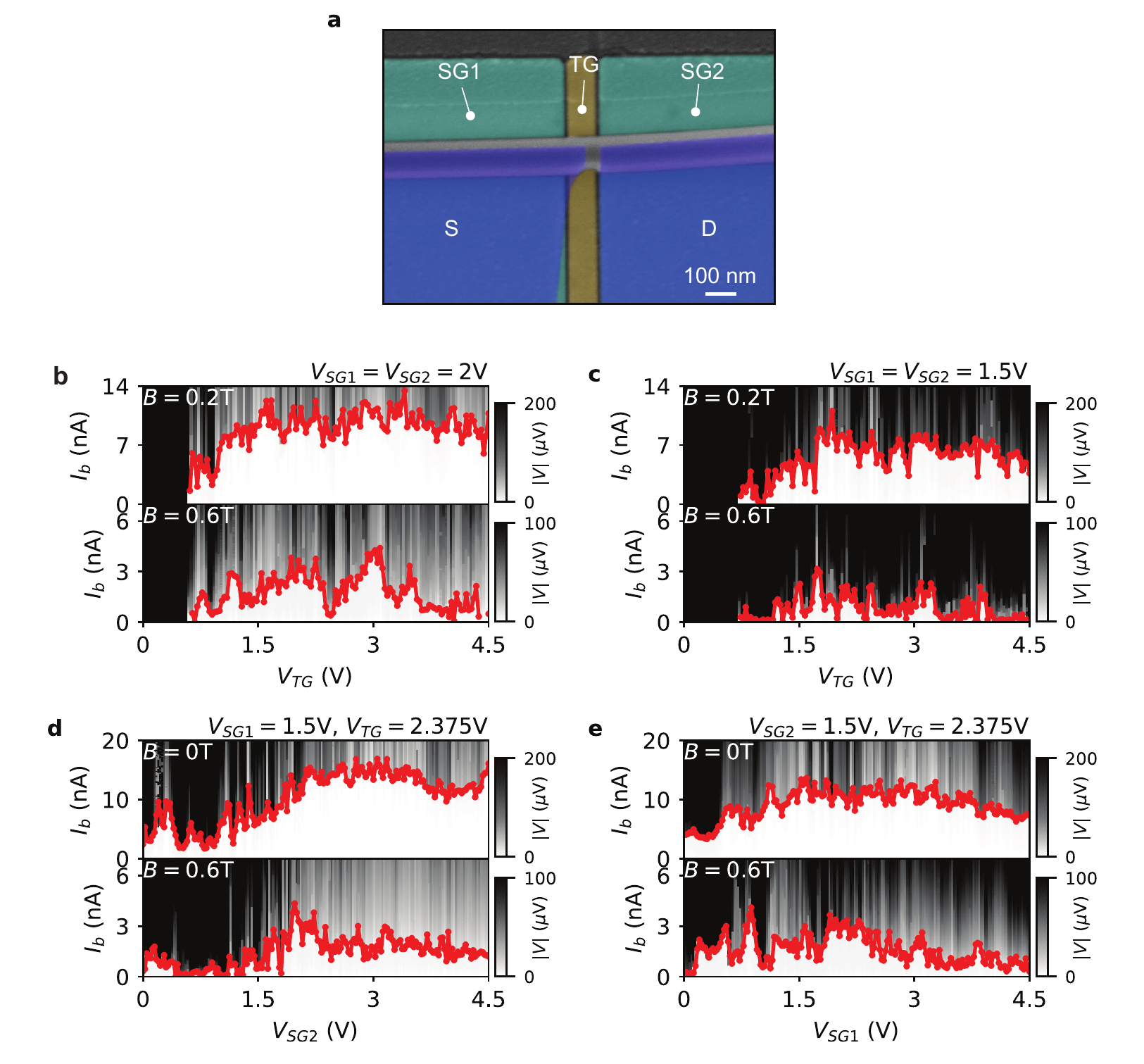}
\caption{\textbf{Effects of local gates on supercurrent in the Josephson junction of Device 10}: Measurements were taken on a single arm of the SQUID, while the other arm was pinched-off. \textbf{(a)} False colour SEM image showing a $40\,\mathrm{nm}$ junction with the three local bottom gates: SG1, TG and SG2 (a zoom-in at the right junction of Fig. 5a; \textbf{(b)-(e)} Dependences of the extracted switching current $I_{sw}$ (red) as a single local bottom gate voltage is swept while the other two local bottom gates and the parallel magnetic field are set as written in the corresponding insets.}\label{fig:S8}
\end{figure}

As an additional measurement, we perform current bias measurements on a single Josephson junction (Device 10) which is one arm of the SQUID (see Fig. \ref{fig:S8}a and the Fabrication section for the details on the device design). The local bottom gates under the nanowire in Device 10 allow for a local tuning of the electro-chemical potential in different sections of the nanowire and can therefore serve to evaluate the effects of the local gating on the supercurrent resilience. 

We perform current bias measurements on Device 10 while the other arm of the SQUID is pinched-off. The three bottom gates - TG and SG1/SG2 - approximately align with the junction and the superconducting leads, as shown in Fig. \ref{fig:S8}a. Two bottom gate voltages $V_{SG1}$ and $V_{SG2}$ mainly tune the nanowire sections covered by the superconductor, while the middle gate voltage $V_{TG}$ mainly tunes the semiconducting junction. In this way the electro-chemical potential in the nanowire can be locally controlled, which is not possible in the global back gate configuration of nanowire JJ devices (Device 1-9) in the main text.

The dielectric used for the local bottom gates is ALD $\mathrm{HfO_2}$ of $\sim 20\,\mathrm{nm}$ thickness. As a comparison, the global back gate of the Device 1-9 utilizes thermal $\mathrm{SiO_2}$ of $\sim 300\,\mathrm{nm}$ thickness. By taking into account the dielectric constant values of $\mathrm{HfO_2}$ and $\mathrm{SiO_2}$ to be $\sim10$ and $\sim4$, respectively, the gating effect of the local bottom-gates is estimated to be at least 30 times larger than that of the global back gate. 

In Fig. \ref{fig:S8}b-e, dependences of the extracted switching current $I_{sw}$ (red) on a single bottom gate voltage are shown, while the other two bottom gates and the parallel magnetic field are fixed. By comparing Fig. \ref{fig:S8}b and Fig. \ref{fig:S8}c, it can be noticed that sweeping just $V_{TG}$ qualitatively resembles the case when the global back-gate is swept (Fig. 1b, Fig. \ref{fig:S2} and Fig. \ref{fig:S7}). When $V_{SG1}$ and $V_{SG2}$ are decreased in Fig. \ref{fig:S8}c in comparison to Fig. \ref{fig:S8}b, a slight decrease in $I_{sw}$ can be observed. This can be attributed to $V_{SG1}$ and $V_{SG2}$ cross-coupling to the junction and effectively reducing its transmission. By looking at Fig. \ref{fig:S8}d and Fig. \ref{fig:S8}e, it can be seen that sweeping a single local bottom gate under the superconducting leads over $4.5\,\mathrm{V}$ does not systematically affect the extracted switching current $I_{sw}$. In some cases, a slight increase in the background value of $I_{sw}$ can be observed as $V_{SG1}$ or $V_{SG2}$ increase over $4.5\,\mathrm{V}$ voltage range. This is also in agreement with $V_{SG1}$ and $V_{SG2}$ cross-coupling to the junction. 

The fluctuations of the switching current magnitude in the single local bottom gate sweeps in Fig. \ref{fig:S8}b-e are comparable to those observed in the global back gate traces in the main text. Therefore, it cannot be determined whether the fluctuations in the back gate sweeps arise from the modulations of the electro-chemical potential of the junction or the nanowire sections under the superconducting leads. Importantly, we observe that applying positive voltage on the single local bottom gate under the superconducting lead does not diminish the semiconductor-superconductor coupling to an extent that the supercurrent of Device 10 is systematically suppressed.

\section{Data selection and reproducibility}

By systematically sweeping the back gate voltage $V_g$ when measuring Device 1-7, we could identify the resilient gate settings $V_{g,res}$, as described in the main text. However, at the initial phase of the study, when measuring the chips from which Device 8-9 originate, the resilience of supercurrent against magnetic field was only examined at $V_g=15\,\mathrm{V}$. Therefore, for these devices the identification of the resilient gate setting $V_{g,res}$ (like those shown in Fig. \ref{fig:S5}) was not performed. Still, we include Device 8-9 in our study as they manifest resilient supercurrent even at $V_g=15\,\mathrm{V}$ which is not necessarily their $V_{g,res}$. Other short junction devices from these chips did not manifest such resilient supercurrent (critical parallel field of $\sim0.7\,\mathrm{T}$ at $V_g=15\,\mathrm{V}$) and long junction devices from these chips showed very poor supercurrent resilience (critical parallel field of $\sim0.4\,\mathrm{T}$ at $V_g=15\,\mathrm{V}$). We do not include these devices in our study as their critical parallel fields at $V_g=15\,\mathrm{V}$ may be significantly smaller in  comparison to their critical fields at the back gate tuned to their $V_{g,res}$ settings.

Importantly, we have never measured any long junction device (with or without back gate tuning) that showed better supercurrent resilience than the long junction devices (Device 6-7) presented in the study.

\bibliography{references}